\documentstyle[12pt,draft,nature_macros]{nature}

\font\large=cmbx10 scaled \magstep3

\newcommand{\equ}[1]{eq.~(\ref{eq:#1})}
\newcommand{\Equ}[1]{Eq.~(\ref{eq:#1})}
\newcommand{\se}[1]{\S\ref{sec:#1}}
\newcommand{\fig}[1]{Fig.~\ref{fig:#1}}
\newcommand{\Fig}[1]{Fig.~\ref{fig:#1}}
\newcommand{\Figs}[1]{Figs.~\ref{fig:#1}}
\newcommand{\be}{\begin{equation}}
\newcommand{\ee}{\end{equation}}

\newcommand{\msun}{M_\odot}
\newcommand{\ifm}[1]{\relax\ifmmode#1\else$\mathsurround=0pt #1$\fi}
\newcommand{\kms}{\ifmmode\,{\rm km}\,{\rm s}^{-1}\else km$\,$s$^{-1}$\fi}
\newcommand{\kpc}{\ifmmode\,{\rm kpc}\else kpc\fi}
\newcommand{\Mpc}{\ifmmode\,{\rm Mpc}\else Mpc\fi}

\newcommand{\ltsima}{$\; \buildrel < \over \sim \;$}
\newcommand{\lsim}{\lower.5ex\hbox{\ltsima}}
\newcommand{\gtsima}{$\; \buildrel > \over \sim \;$}
\newcommand{\gsim}{\lower.5ex\hbox{\gtsima}}
\newcommand{\prop}{\propto}

\def\no{\noindent}
\def\Rv{R_{\rm v}}
\def\Mv{M_{\rm v}}
\def\Vv{V_{\rm v}}
\def\fb{f_{\rm b}}

\def\sy{\,M_\odot\, {\rm yr}^{-1}}
\def\Mpc3{\,{\rm Mpc}^{-3}}

\def\ifm#1{\relax\ifmmode#1\else$\mathsurround=0pt #1$\fi}
\def\kms{\,{\rm km\,s\ifm{^{-1}}}}
\def\hmpc{\,h\ifm{^{-1}}{\rm Mpc}}
\def\hkpc{\,h\ifm{^{-1}}{\rm kpc}}
\def\Gyr{\,\rm Gyr}
\def\Myr{\,\rm Myr}

\newcommand{\ad}[1]{}

\begin{document}

\title{Cold streams in early massive hot haloes\\
as the main mode of galaxy formation}

\author{A. Dekel$^{1}$, Y. Birnboim$^{1}$, 
G. Engel$^{1}$, J. Freundlich$^{1,2}$, T. Goerdt$^1$, 
M. Mumcuoglu$^{1}$, E. Neistein$^1$, C. Pichon$^3$, R. Teyssier$^{4,5}$, 
\& E. Zinger$^1$ 
\institute{$^1$Racah Institute of Physics, 
               The Hebrew University, Jerusalem 91904, Israel\\
$^2$Department de Physique, Ecole Normale Superieure, 24 rue Lhomond,
           75231 Paris cedex 05, France\\ 
$^3$Institut d'Astrophysique de Paris and UPMC, 
98bis Boulevard Arago, Paris 75014, France\\
$^4$Institut de Recherches sur les lois Fondamentales de l'Univers, DSM, 
l'Orme des Merisiers, 91198 Gif-sur-Yvette, France\\
$^5$Institute for Theoretical Physics, University of Zurich, CH-8057 Zurich,
Switzerland
}
}

\dates{}{}
\mainauthor{Dekel et al.}
\headertitle{Galaxy formation by cold streams}

\summary{The massive galaxies in the young Universe, ten billion years ago,
formed stars at surprising intensities\cite{genzel06,chapman04}.
Although this is commonly
attributed to violent mergers, the properties of many of these
galaxies are incompatible with such events, showing gas-rich,
clumpy, extended rotating disks not dominated by
spheroids\cite{genzel06,forster06,genzel08}.
Cosmological simulations\cite{ocvirk08} and clustering
theory\cite{neistein06,neistein08b} are used to
explore how these galaxies acquired their gas. Here we report that
they are stream-fed galaxies, formed from steady, narrow, cold gas
streams that penetrate the shock-heated media of massive dark
matter haloes\cite{db06,keres05}. A comparison with the observed abundance of
star-forming galaxies implies that most of the input gas must
rapidly convert to stars. One-third of the stream mass is in gas
clumps leading to mergers of mass ratio greater than 1:10, and
the rest is in smoother flows. With a merger duy cycle of 0.1,
three-quarters of the galaxies forming stars at a given rate are fed
by smooth streams. The rarer, submillimetre galaxies that form
stars even more intensely\cite{chapman04,wall08,tacconi08}
are largely merger-induced starbursts.
Unlike destructive mergers, the streams are likely to keep
the rotating disk configuration intact, although turbulent and
broken into giant star-forming clumps that merge into a central
spheroid\cite{noguchi99,elmegreen08,genzel08,dekel09}.
This stream-driven scenario for the formation of disks and spheroids is
an alternative to the merger picture.}
\maketitle

 
\noindent{\bf Star-Formation Rate versus Halo Growth Rate}

\noindent
It appears that the most effective star formers in the Universe were galaxies
of stellar and gas mass of $\sim\!10^{11}\msun$ at redshifts 
$z\!=\!2\!-\!3$, when the Universe was
$\sim\!3\Gyr$ old.  
The common cases\cite{genzel06,forster06}
show star-formation rates (SFR) of $100\!-\!200\sy$.
These include UV-selected galaxies termed BX/BM\cite{adelberger04}
and rest-frame optically selected galaxies termed sBzK\cite{daddi04},
to be referred to collectively as ``Star-Forming Galaxies" (SFGs). 
Their SFRs are much higher than the $4\sy$ in today's Milky Way,
while their masses and dynamical times are comparable.
The comoving space density of SFGs is $n\!\simeq\!2\!\times\!10^{-4}\Mpc3$,
implying within the standard cosmology (termed $\Lambda$CDM) that they 
reside in dark-matter haloes of masses $\lsim\!3.5\!\times\!10^{12}\msun$. 
The most extreme star formers are dusty 
Sub-Millimeter Galaxies (SMG)\cite{tacconi08,wall08},
with SFRs of up to $\sim\!1,000\sy$ and $n\!\simeq\!2\!\times\!10^{-5}\Mpc3$.
Whereas most SMGs could be starbursts induced by major mergers,
the kinematics of the SFGs indicate extended, clumpy, thick
rotating disks that are incompatible with the expected compact or
highly perturbed kinematics of ongoing 
mergers\cite{forster06,genzel06,bouche07,genzel08}.
The puzzle is how massive galaxies form most of their stars so
efficiently at early times through a process other than a major merger. 
A necessary condition is a steady, rapid gas supply into massive disks.

It is first necessary to verify that the required rate of gas supply is
compatible with the cosmological growth rate of dark matter haloes.
The average growth rate of halo mass, $\Mv$, through mergers and smooth
accretion, is derived\cite{neistein06} based on the EPS theory of 
gravitational clustering\cite{lacey93} (Supplementary Information, SI, \S 1),
or from cosmological simulations\cite{neistein08a,genel08}.  
For $\Lambda$CDM, the corresponding growth rate of the baryonic
component is approximately  
\be
\dot{M} \simeq 6.6\, M_{12}^{1.15}\, (1+z)^{2.25}\, f_{.165}\, \sy \ ,
\label{eq:mdot_vir_2}
\ee
where $M_{12}\equiv\Mv/10^{12}\msun$,
and $f_{.165}$ is the baryonic fraction in the haloes
in units of the cosmological value, $\fb\!=\!0.165$.
Thus, at $z\!=\!2.2$, the baryonic growth rate of haloes
of $2\!\times\!10^{12}\msun$ is $\dot{M}\!\simeq\!200\sy$,
sufficient for fueling the SFR in SFGs.
However, the margin by which this is sufficient is not large,
implying that (1) the incoming material must be mostly gaseous,
(2) the cold gas must efficiently penetrate into the inner halo,
and (3) the SFR must closely follow the gas supply rate.

\begin{figure}
\vskip 18.9cm
{\includegraphics{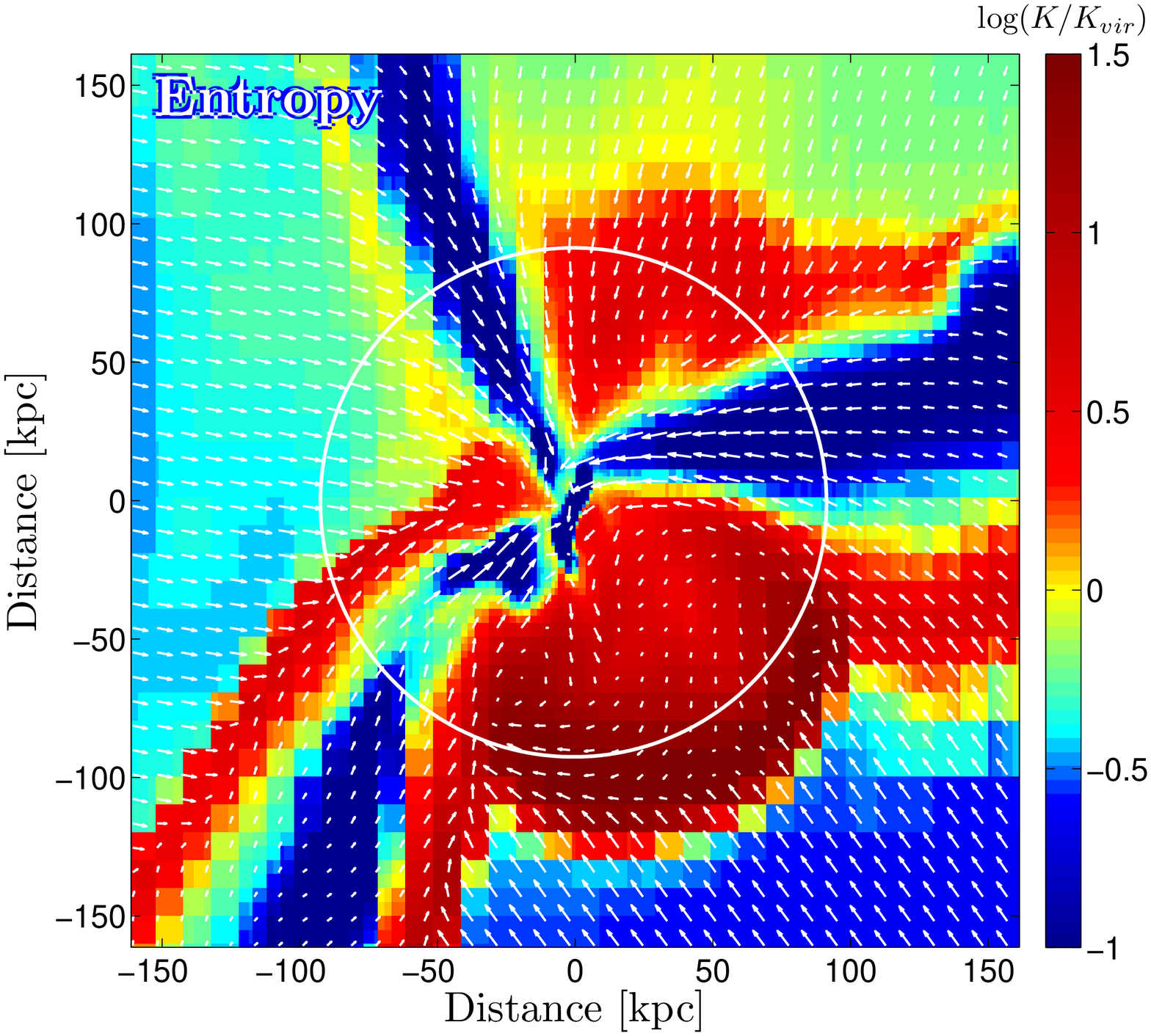}}
{\includegraphics{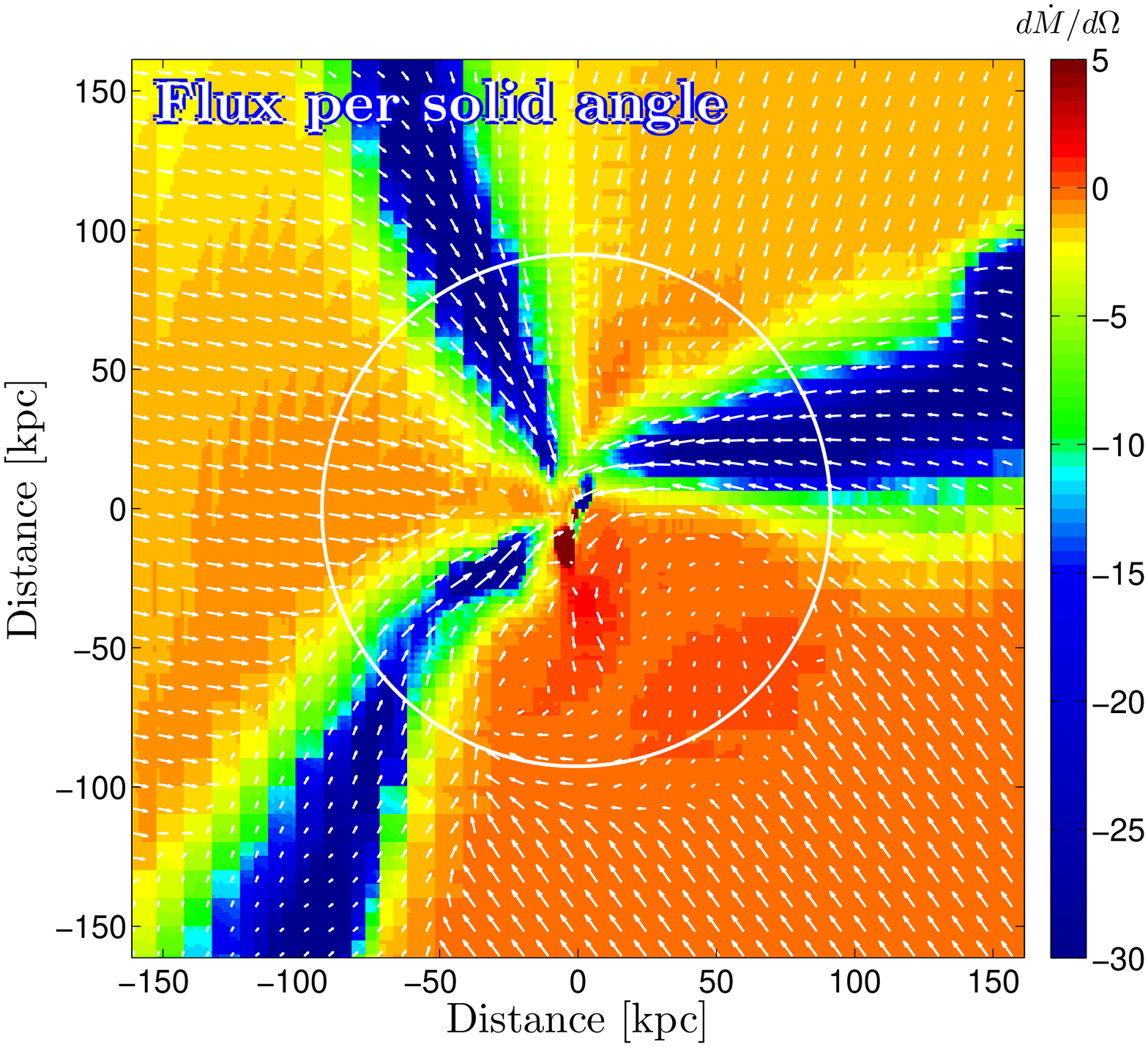}}
\vskip 0.2cm
\caption{\small 
Entropy, velocity and inward flux of cold streams pouring through hot haloes.
The maps refer
to a thin slice through one of our fiducial galaxies of $\Mv\!=\!10^{12}\msun$
at $z\!=\!2.5$.  The arrows describe the velocity field, scaled such that the 
distance between the tails is $260\kms$.  The circle marks the halo virial 
radius $\Rv$.  The {\bf entropy}, $\log K\!=\!\log (T/\rho^{2/3})$, in units of
the virial quantities, highlights (in red) the high-entropy medium filling the 
halo out to the virial shock outside $\Rv$.  It exhibits (in blue) three,  
radial, low-entropy streams that penetrate into the inner disk, seen edge-on.
The radial {\bf flux} per solid angle is $\dot{m}\!=\!r^2 \rho\, v_r$, in 
$\!\sy{\rm rad}^{-2}$, where $\rho$ is the gas density and $v_r$ the radial
velocity.  
}
\label{fig:1}
\end{figure}

\medskip\noindent{\bf Penetrating Cold Narrow Streams}

\noindent
The deep penetration is not a
trivial matter, given that the halo masses of $\Mv\!>\!10^{12}\msun$ are
above the threshold for virial shock heating\cite{bd03,binney04,keres05,db06},
$M_{\rm shock}\!\lsim\!10^{12}\msun$.
Such haloes are encompassed by a stable shock near their outer radius, $\Rv$,
inside which gravity and thermal energy are in virial equilibrium.
Gas falling in through the shock is expected to heat up to the virial
temperature and stall in quasi-static equilibrium before it cools and
descends into the inner galaxy\cite{bdn07}.  However,
at $z \geq 2$, these hot haloes are penetrated by cold
streams\cite{keres05,db06}.
Dekel \& Birnboim\cite{db06} pointed out that because early haloes with
$\Mv\!>\!M_{\rm shock}$ populate the massive tail of the distribution,
they are fed by dark-matter filaments from the cosmic web that
are narrow compared to $\Rv$ and denser than the mean within the halo.
The enhanced density of the gas along these filaments makes the flows along 
them unstopable; in particular, they 
cool before they develop the pressure to support a shock,
and thus avoid the shock heating (SI, \S 2).

\begin{figure}[h]
\vskip 9.0cm
{\includegraphics{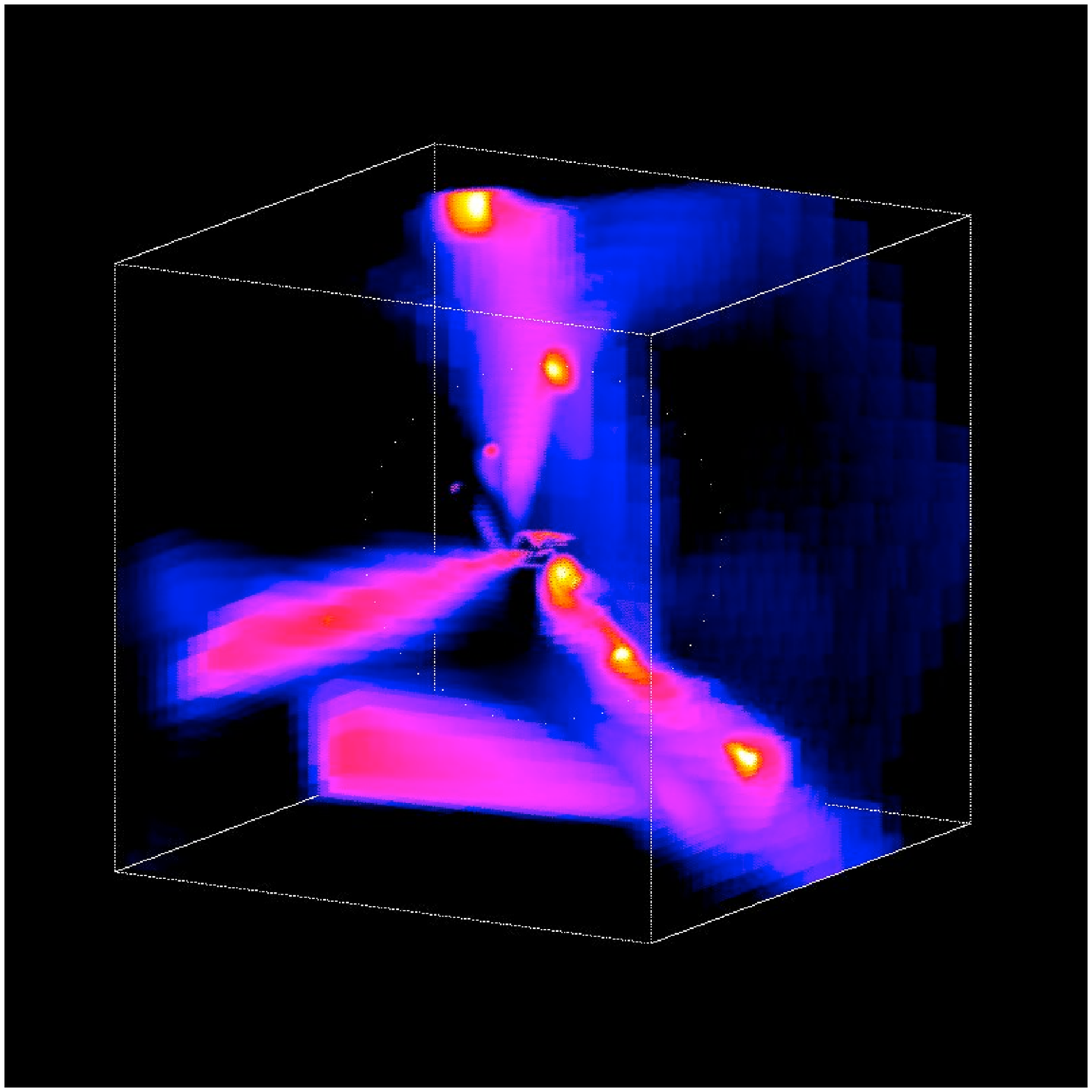}}
\vskip 0.2cm
\caption{\small
Streams in three dimensions. The map shows radial flux for the galaxy of
Fig.~1 in a box of side length $320\kpc$. 
The colours refer to inflow rate per solid
angle of point-like tracers at the centers of cubic-grid cells. 
The dotted circle marks the halo virial radius.
The appearance of three fairly radial streams seems to be generic in
massive haloes at high redshift --- a feature of the cosmic web that deserves
an explanation.
Two of the streams show gas clumps of mass on the order of one-tenth of the
central galaxy, but most of the stream mass is smoother 
(SI, Fig.~10).
The $\gsim\!10^{10}\msun$ clumps, which involve about 
one-third of 
the incoming mass, are also gas rich --- in the current simulation only 30\% of
their baryons turn into starts before they merge with the central galaxy.
}
\label{fig:2}
\end{figure}

To investigate the penetration of cold streams, we study the
way gas feeds massive high-$z$ galaxies in the cosmological
MareNostrum simulation ---
an adaptive-mesh hydrodynamical simulation in a comoving box of side length
$71\,{\rm Mpc}$ and a resolution of $1.4\kpc$ at the galaxy centers (SI \S 3).
The gas maps in \Figs{1} \& \ref{fig:2}
demonstrate how the shock-heated, high-entropy, low-flux medium
that fills most of the halo
is penetrated by three narrow, high-flux streams of low-entropy gas
(SI, Figs.~7-10).
The flux map
demonstrates that more than 90\% of the inflow is channeled
through the streams (blue), at a rate that remains roughly the same at all
radii. This rate is several times higher than the spherical average outside the
virial sphere, $\dot{m}_{\rm vir}\!\simeq\!8\sy {\rm rad}^{-2}$ by
\equ{mdot_vir_2}.
The opening angle of a typical stream at $\Rv$ is $20^\circ\!-\!30^\circ$,
so the streams cover a total angular area of $\sim\!0.4\,{\rm rad}^2$, namely
a few percent of the sphere.
When viewed from a given direction, the column density of cold gas below
$10^5$K is above $10^{20} {\rm cm}^{-2}$ for 25\% of the area within the virial
radius.
While the pictures show the inner disk, the disk width is not resolved, so the
associated phenomena such as shocks, star formation and feedback are treated in
an approximate way only.

\begin{figure}[h]
\vskip 8.7cm
{\includegraphics{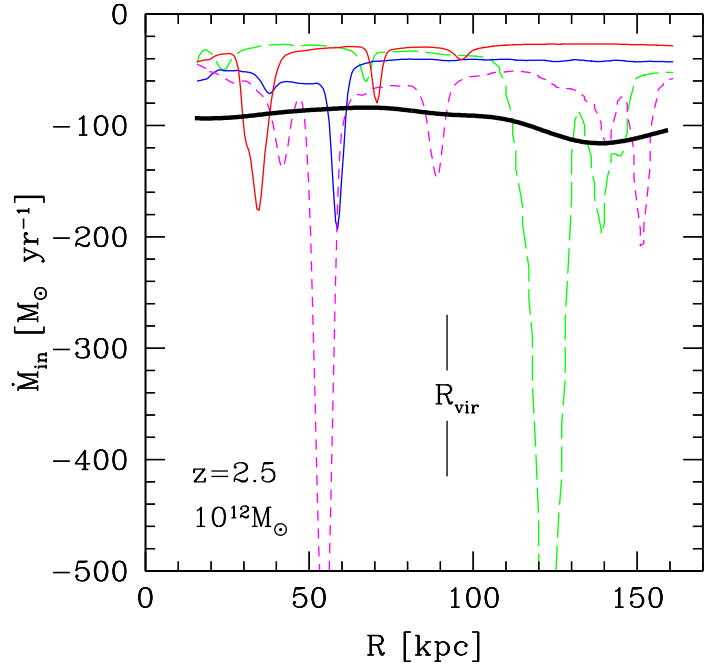}}
\vskip 0.2cm
\caption{\small 
Accretion profiles $\dot{M}(r)$.
Shown is the gas inflow rate through spherical shells of radius
$r$, from the disk vicinity to almost twice the halo virial radius,
obtained by integrating $r^2 \rho\,v_r$ over the whole shell.
The thick black curve is the
average over the simulated galaxies of the fiducial case,
$\Mv\!\simeq\!10^{12}\msun$ at $z\!=\!2.5$. It shows deep penetration at a
roughly constant rate $\sim\!100 \sy$, consistent with the virial growth rate
predicted by \equ{mdot_vir_2}.  Apparently, the inflow rate does decay while
traveling
through the halo, but this decay is roughly compensated by the higher
cosmological inflow rate when that gas entered the halo, \equ{mdot_vir_2},
leading to the apparent constancy of accretion rate with radius.
The coloured curves refer to four
representative galaxies, two showing clumps with $\mu \gsim 0.1$ and two with
smoother flows involving only mini-minor clumps of $\mu < 0.1$.  Clumps with
$\mu\!\gsim\!0.3$ appear within $2\Rv$ about once in ten galaxies; that is,
major mergers are infrequent 
(SI, Fig.~11).
The $\dot{M}(r)$ profiles serve for extracting the conditional probability
distribution $P(\dot{M}|\Mv)$, leading to the abundance $n(>\dot{M})$
(SI, Fig.~12).
}
\label{fig:3}
\end{figure}

The penetration is evaluated from the
profiles of gas inflow rate, $\dot{M}(r)$, through shells of radius $r$,
\Fig{3} 
(SI, Fig.~11).
The average profile reveals that the flow rate remains constant
from well outside $\Rv\!\sim\!90\kpc$ to the disk inside $r\!\sim\!15 \kpc$.


\medskip\noindent{\bf Abundance of Gas Inflow Rates}

\begin{figure}[h]
\vskip 8.7cm
{\includegraphics{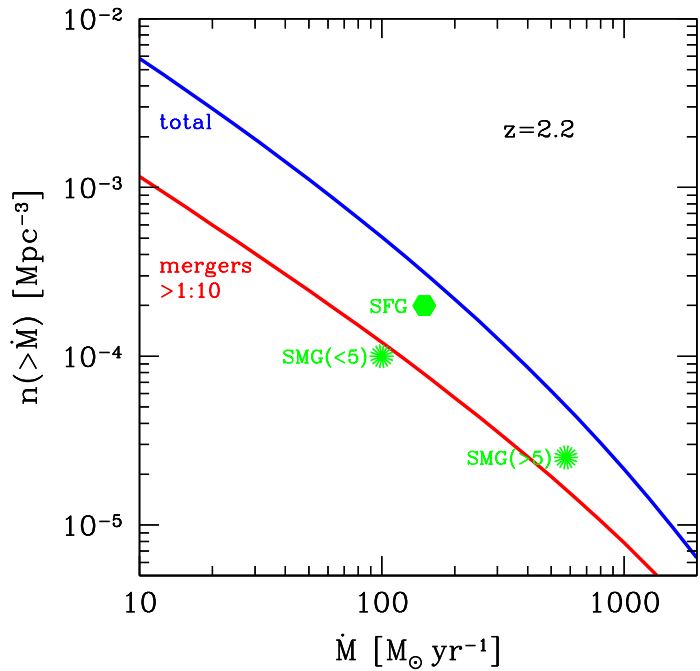}}
\vskip 0.2cm
\caption{\small 
Abundance of galaxies as a function of gas inflow rate, $n(>\dot{M})$.
Shown is the
comoving number density, $n$, of galaxies with inflow rate higher than 
$\dot{M}$
at $z\!=\!2.2$, as predicted from our analysis of the cosmological simulation.
The upper curve refers to total inflow.  It shows that galaxies with 
$\dot{M}\!>\!150\sy$ are expected at a comoving number density 
$n\!\sim\!3\!\times\!10^{-4}\Mpc3$ (similar to estimates in other 
simulations\cite{finlator06,nagamine08}). Fluxes as high as
$\dot{M}\!>\!500\sy$
are anticipated at $n\!\sim\!6\!\times\!10^{-5}\Mpc3$.
The lower curve is similar, but limited to gas input by $\mu\!>\!0.1$ mergers.
The symbols represent the vicinity of where the observed massive star-forming
galaxies can be located once their observed SFR is identified with $\dot{M}$.
The sBzK/BX/BM galaxies are marked SFG\cite{tacconi08}.  The SMGs respectively
brighter and
fainter than 5 mJy are marked accordingly\cite{tacconi08,wall08}.
We see that the overall gas inflow rate is sufficient for the observed SFR, but
the small margin implies that the SFR must closely follow the rate of 
gas supply. 
Most of the massive star formers at a given SFR 
are expected to be observed while being fed by 
smooth flows rather than undergoing mergers.
By studying the contribution of different halo masses to the abundance
$n(>\dot{M})$, we learn that the high-SFR SFGs and SMGs are associated with
haloes of mass $10^{12}-10^{13}\msun$ 
(SI, Fig.~13).
}
\label{fig:4}
\end{figure}

\noindent
To relate the feeding by streams to the observed abundance of galaxies as a 
function of SFR, we use the MareNostrum inflow-rate profiles to evaluate 
$n(>\!\dot{M})$, 
the comoving number density of galaxies with an inflow rate $>\!\dot{M}$.
We first extract  
the conditional probability distribution $P(\dot{M}|\Mv)$
by sampling the $\dot{M}(r)$ profiles uniformly in $r$, 
using the fact  
that the velocity along the streams is roughly constant (SI \S\S5,6).
This is convolved with the halo mass function\cite{sheth02}, $n(\Mv)$,
to give   
\be
n(\dot{M}) = \int_0^\infty P(\dot{M}|\Mv)\, n(\Mv)\, d\Mv \ .
\label{eq:nmdot}
\ee
The desired cumulative abundance $n(>\!\dot{M})$, obtained by integration
over the inflow rates from $\dot{M}$ to infinity, 
is shown at $z\!=\!2.2$ in \fig{4}. 
Assuming that the SFR equals $\dot{M}$, 
the curve referring to $\dot{M}$ lies safely
above the observed values, marked by the symbols,
indicating that the gas input rate is sufficient to explain the SFR.
However, $\dot{M}$ and the SFR are allowed to differ only by
a factor of $\sim\!2$, confirming our suspicion that
the SFR must closely follow the gas-input rate. 
Because at $z\!\sim\!2.2$ the star-forming galaxies constitute only a 
fraction of the
observed $\sim\!10^{11}\msun$ galaxies\cite{kriek06,dokkum08}, 
the requirement for a SFR SFR almost as great as $\dot{M}$, based on \fig{4},
becomes even stronger.


\medskip\noindent{\bf Smooth Flows versus Mergers}

\noindent
By analysing the clumpiness of the gas streams, using the sharp peaks of
inflow in the $\dot{M}(r)$ profiles,
we address the role of mergers versus smooth flows.
We evaluate each clump mass by integrating
$M_{\rm clump}\!=\!\int \dot{M}(r)\, dr/v_r(r)$ across the peak,
and estimate a mass ratio for the expected merger
by $\mu\!=\!M_{\rm clump}/(\fb \Mv)$, ignoring
further mass loss in the clump on its way in
and deviations of the galaxy baryon fraction from $\fb$.
We use the term ``merger" to describe any major or minor merger of 
$\mu\!\geq\!0.1$, as distinct from 
``smooth" flows, which include ``mini-minor'' mergers with $\mu\!<\!0.1$.
We find that about one-third of the mass is flowing in as mergers
and the rest as smoother flows.
However, the central galaxy is fed by a clump of $\mu\!\geq\!0.1$
during less than 10\% of the time; that is,  
the duty cycle for mergers is $\eta\!\lsim\!0.1$.
A similar estimate is obtained using EPS merger rates\cite{neistein08b}
and starburst durations of $\sim\!50 \Myr$ at $z\!=\!2.5$
from simulations\cite{cox08} (SI, \S 5).

From the difference between the two curves of \fig{4}
we learn that only a quarter of the galaxies with a given $\dot{M}$
are to be seen during a merger.  The fact that the SFGs 
lie well above the merger curve even if the SFR is $\sim\!\dot{M}$
indicates that in most of them the star formation is driven by smooth streams.
Thus, ``SFG" could also stand for ``Stream-Fed Galaxies". 
This may explain why these galaxies maintain an extended, thick disk
while doubling their mass over a halo crossing time\cite{genzel08}.
On the other hand, if the SFR is $\sim\!\dot{M}$,  
we learn from \fig{4} that about half of the bright SMGs
and most of the fainter SMGs lie below the merger curve and are therefore
consistent with being merger-induced starbursts\cite{tacconi08}.


\medskip\noindent{\bf Conclusion and Discussion}

\noindent
We obtain that Stream-Fed Galaxies of baryonic mass $\sim\!10^{11}\msun$ at
$z\!\sim\!2.5$ were the most productive star formers in the Universe.
An integration of $\dot{M}$ over halo mass and time reveals that
most of the stars in the Universe were formed in Stream-Fed Galaxies,
within haloes $>\!2\times 10^{11}\msun$ at $1.5<z<4$.
The constraints on the overall SFR density at these epochs\cite{hopkins04}
imply that SFR has been suppressed in smaller galaxies, e.g., by
photoionization and stellar feedback\cite{ds86,dw03,db06}.
The early presence of low-SFR galaxies\cite{kriek06,dokkum08}
requires quenching of SFR also at the massive end,
perhaps due to gravitational heating by destructive streams\cite{db08}.

The streams are likely also to be responsible for compact
spheroids, as an alternative to mergers\cite{robertson08}
and the associated heating by expanding shocks\cite{bdn07,db08}.
Using \equ{mdot_vir_2}, we find that at $z\!\geq\!2$
the streams can maintain both the high gas fraction and the turbulence
necessary for the disk to breakup into giant clumps by gravitational 
instability, with dispersion-to-rotation ratio $\sigma/V\!\sim\!0.25$,
as observed\cite{genzel08}.
The clumps migrate inward and dissipatively merge into a
spheroid\cite{noguchi99,elmegreen08}.
The stream carrying the
largest coherent flux with an impact parameter of a few kiloparsecs 
determines the
disk's spin and orientation, and the stream clumps perturb it.
The incoming clumps and the growing spheroid can eventually stabilize 
the disk and suppress star formation.
We can thus associate the streams with the main mode of galaxy
and star formation occurring in massive haloes at $z\!\sim\!2\!-\!3$;
the streams that create the disks also make them fragment into giant clumps
that serve both as the sites of efficient star formation and the progenitors
of the central spheroid, which in turn helps the
streams to quench star formation.

Although wet mergers may grow secondary disks31, they are not as
frequent as the observed SFGs (\fig{4}), these disks are neither gas rich
nor clumpy enough, and, unlike most SFGs, they are dominated by
stellar spheroids.

The cold streams should be detectable by absorption or emission.
For external background sources, our simulation predicts
that haloes with $\Mv\!\sim\!10^{12}\msun$ at $z\!\sim\!2.5$
should contain gas at temperature $<\!10^5$K with column densities 
$>\!10^{20}\,{\rm cm}^{-2}$
covering $\sim\!25\%$ of the area at radii between $20$ and $100\kpc$,
with coherent velocities $\lsim\!200\kms$.
Sources at the central galaxies should show absorption by
the radial streams in $\sim\!5\%$ of the galaxies,
flowing in at $\gsim\!200\kms$, with column densities
$\sim\!10^{21}\,{\rm cm}^{-2}$ 
(SI, Figs.~7-9).


\addtolength{\baselineskip}{-0.05\baselineskip}

\bibliographystyle{natureedo}
\bibliography{flows_nat}


\vskip 0.5cm
\noindent{\bf Acknowledgments}
We acknowledge stimulating diskussions with N. Bouche, S.M. Faber, R. Genzel,
D. Koo, A. Kravtsov, A. Pope, J.R. Primack, J. Prochaska, 
A. Sternberg \& J. Wall.
This research has been supported by the France-Israel Teamwork in Sciences,
the German-Israel Science Foundation, the Israel Science Foundation,
a NASA Theory Program at UCSC, and a Minerva fellowship (TG).
We thank the computer resources and technical support by the
Barcelona Centro Nacional de Supercomputacion.
The simulation is part of the Horizon collaboration.

\vskip 0.5cm
\noindent{\bf Author Information}
Correspondence and requests for materials should be
addressed to A.D. (dekel@phys.huji.ac.il).




\newcommand{\figs}[1]{Figs.~\ref{fig:#1}}

\def\msh{M_{\rm shock}}
\def\mps{M_{*}}
\def\mst{M_{\rm stream}}
\def\zc{z_{\rm crit}}


\vfill\eject
\centerline{\bf SUPPLEMENTARY INFORMATION}


\bigskip\no
This is an extension of the Letter to Nature,
aimed at providing further details, in support of the results reported 
in the main body of the Letter.  

\section{Halo growth by EPS}
\label{sec:eps}

Neistein et al.\cite{neistein06} used the EPS\cite{lacey93} theory of
cosmological clustering into spherical haloes in virial equilibrium
to derive a robust approximation for the average growth rate of
halo virial mass $\Mv$, 
\be  
d\ln \Mv/d\omega = -(2/\pi)^{1/2} [\sigma^2(\Mv/q)-\sigma^2(\Mv)]^{-1/2} ,\quad
\omega \equiv \delta_{\rm c}/D(t) \ .
\label{eq:mdot_vir_1}
\ee
The time variable $\omega$, which makes the expression time invariant,
is inversely proportional to $D(t)$,
the linear growth rate of density fluctuations at time $t$ in the assumed
cosmology, with $\delta_{\rm c}\!\simeq\! 1.68$.  
The power spectrum of initial density fluctuations enters via the variance
$\sigma^2(\Mv)$. The constant $q$ is $2.2$ with an uncertainty of $\pm0.1$
intrinsic to the EPS theory.
\Equ{mdot_vir_1} has been confirmed to resemble the assembly rate in
cosmological $N$-body simulations\cite{neistein08a}.

For the $\Lambda$CDM cosmology\cite{komatsu08_wmap5},
a flat Universe with 72\% dark energy, mass dominated by cold dark matter,
and fluctuation normalization parameter $\sigma_8\!=\!0.8$, 
the corresponding growth rate of the baryonic component
is well fitted by the practical formula\cite{neistein06}
\be
\dot{M} \simeq 6.6\, M_{12}^{1.15}\, (1+z)^{2.25}\, f_{.165}\, \sy \ ,
\label{eq:mdot_vir_2}
\ee
where $M_{12}\equiv\Mv/10^{12}\msun$,
and $f_{.165}$ is the baryonic fraction in the matter assembled into haloes
in units of the cosmological value $f_{\rm b}\!=\!0.165$.

\section{On the origin of narrow streams}
\label{sec:origin}

\begin{figure}[h]
\vskip 8.4cm
{\includegraphics{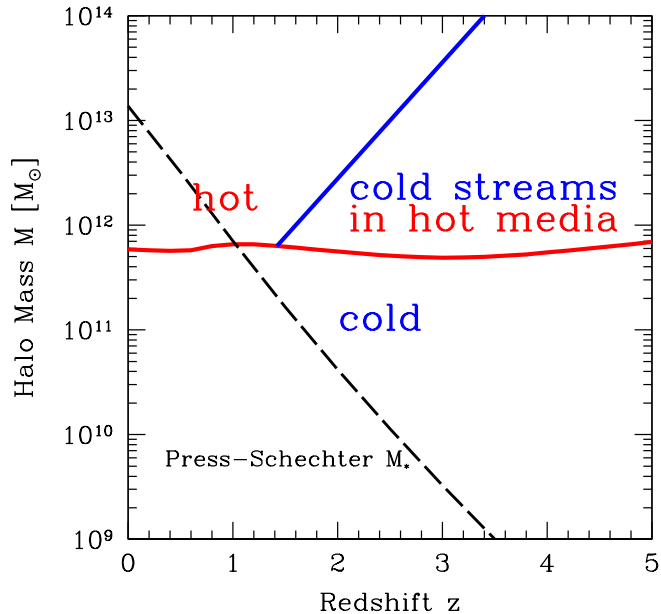}}
\caption{Analytic prediction for the regimes dominated by cold flows and
shock-heated medium in the plane of halo mass and redshift, based on
Fig.~7 of DB06.
The nearly horizontal curve marks the robust threshold mass for a stable
shock based on spherical infall analysis, $\msh(z)$.
Below this curve the flows are predicted to be predominantly cold 
and above it a shock-heated medium is expected to extend out to the halo virial
radius.
The inclined solid curve is the conjectured upper limit for cold streams,
valid at redshifts higher than $z_{\rm crit}\!\sim\!2$.
The hot medium in haloes of $\Mv\!>\!\msh$ at $z\!>\!z_{\rm crit}$
is predicted to host penetrating cold streams, while
haloes of a similar mass at $z\!<\!z_{\rm crit}$ are expected to be all hot,
shutting off most of the gas supply to the inner galaxy. 
Also shown is the characteristic Press-Schechter halo mass $\mps(z)$;
it is much smaller than $\msh$ at $z\!>\!2$.
}
\label{fig:coolzmf}
\end{figure}

Dekel \& Birnboim 2006\cite{db06} (hereafter DB06),
following the simulations by Keres et al. (2005, \S 6.2) and 
their own analysis of the simulations by A. Kravtsov,
pointed out 
that at redshifts higher than $z_{\rm crit}\sim 2$, narrow
cold streams penetrate deep into the dark-matter haloes even when the haloes
are more massive than the shock-heating scale,
$\msh \lsim 10^{12}\msun$, 
and proposed a possible explanation for this phenomenon. 
This prediction is summarized in \fig{coolzmf}.

The critical condition for a stable virial shock is that the 
radiative cooling rate behind the shock is slower than the compression rate, 
$t_{\rm cool}^{-1}\!<\!t_{\rm comp}^{-1}$, allowing the buildup of pressure
support behind the shock against global gravitational collapse\cite{bd03}.
Based on a spherical analysis, DB06 found that a virial shock should exist 
in dark-matter haloes above a threshold mass $\msh\!\lsim\!10^{12}\msun$
that is rather constant in time, at an actual value that is sensitive to the  
metallicity of the gas in the halo.  
The existence of such a threshold mass and its value as a function of redshift
have been confirmed by the analysis of cosmological
simulations\cite{keres05,db06,bdn07,ocvirk08}.
However, at high redshifts, even above the threshold mass,
a shock is not expected to develop along narrow, cold, radial streams 
that penetrate through the halo, because the cooling there is more efficient 
than in the surrounding halo.

The appearance of intense streams at high $z$, as opposed to their absence
at low $z$,
is likely to 
reflect the interplay between the shock-heating scale
and the independent characteristic scale of nonlinear clustering,
i.e., the Press-Schechter\cite{press74} mass $\mps$ that
corresponds to the typical dark-matter haloes forming at a given epoch.
The key difference between the two epochs is that the rapid growth of $\mps$
with time, as seen in \fig{coolzmf}, makes
$\msh\!\gg\!\mps$ at $z\!>\!2$ while $\msh\!\sim\!\mps$ at lower redshifts.

\begin{figure}[h]
\vskip 10.6cm
{\includegraphics{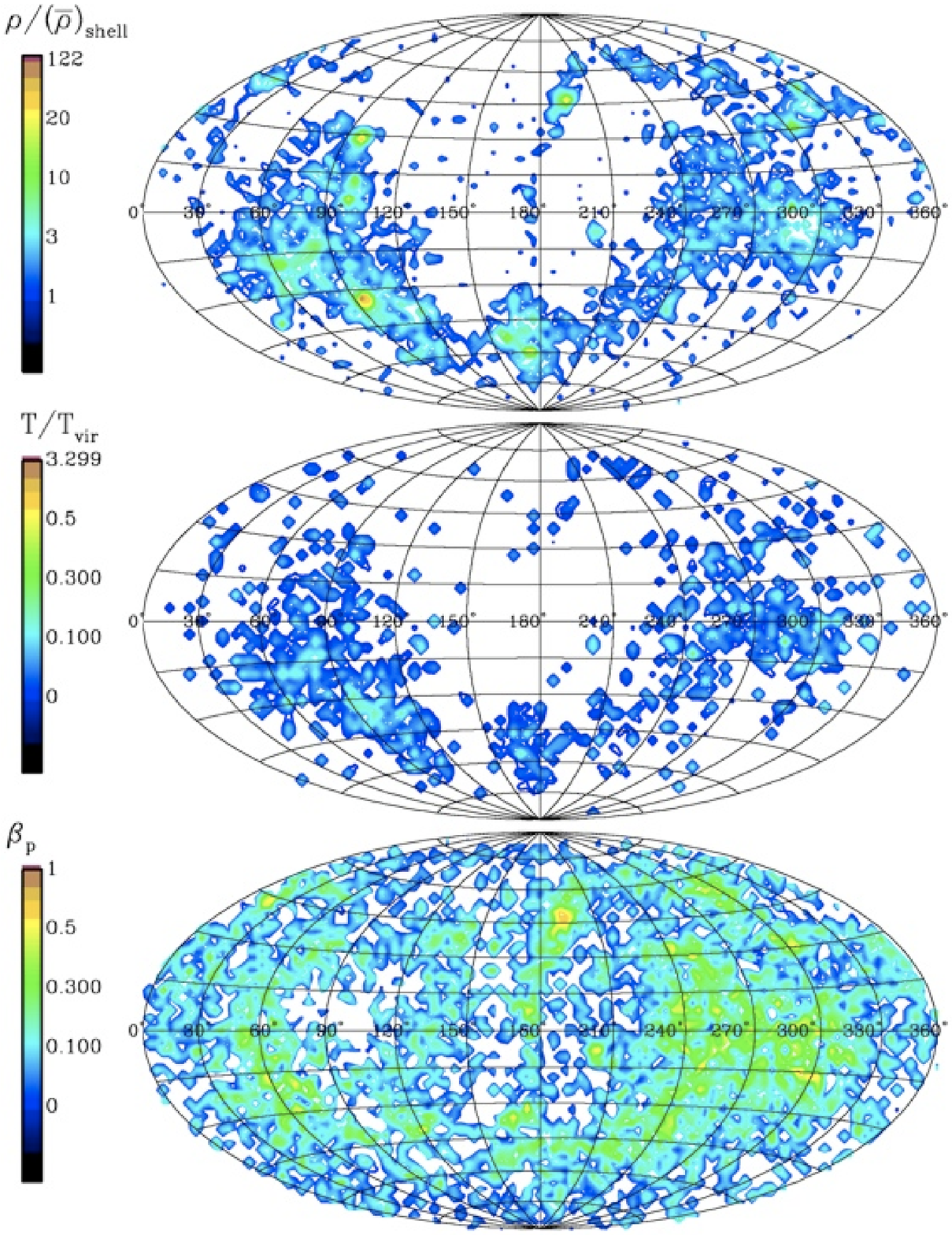}
 \includegraphics{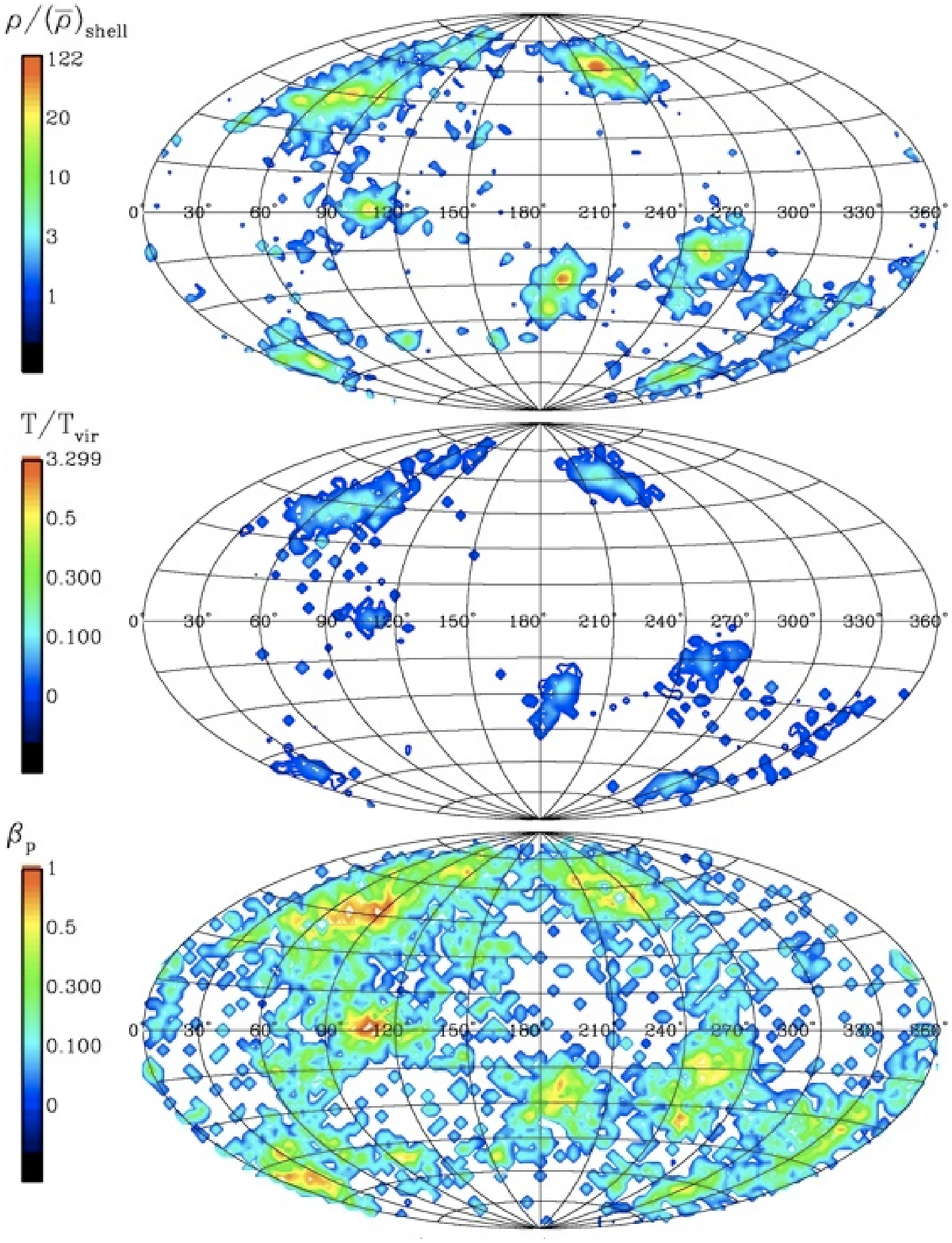}
}
\vskip 0.2cm
\caption{The pattern of dark-matter inflow 
in a shell $(1\!-\!3)\Rv$ outside two sample
haloes from a cosmological N-body simulation at $z=0$
(based on P. Seleson \& A. Dekel, in preparation).
{\bf Left:} a typical halo with $\Mv\!\sim\!M_*$.
{\bf Right:} a rare halo with $\Mv\!\gg\!M_*$. 
In terms of the different ways by which these two haloes are fed by
dark-matter,
they correspond to two haloes of the same mass $\Mv\!\sim\!10^{12}\msun$,
but at $z\!\sim\!0$ and $z\!\sim\!2-3$ respectively.
{\bf Top:} dark-matter density contrast about the mean density in the shell. 
{\bf Middle:} particle velocity dispersion (loosely termed ``temperature"),
in terms of the virial value. 
{\bf Bottom:} infall velocity, represented by the anisotropy parameter $\beta$,
where $\beta=0$ corresponds to isotropic velocities and $\beta=1$ to pure
radial motions.
We see that the typical halo resides inside a broad filament so it is
practically fed by wide-angle diffuse accretion.
On the other hand, the rare halo is fed by narrow, dense, 
radially in-flowing filaments. 
}
\label{fig:aitoff}
\end{figure}

Cosmological $N$-body simulations\cite{keres05,springel05_MR} reveal that
while the rare dark-matter haloes of $\Mv\!\gg\!\mps$ tend to form at the
nodes of intersection of a few filaments of the cosmic web, the typical haloes
of $\Mv\!\sim\!\mps$ tend to reside inside such filaments.
Since the filament width is comparable to the typical halo size
$R_*\!\propto\!\mps^{1/3}$ and seems not to vary much with position
along the filament,
one expects the rare haloes to be fed by a few streams that are narrow compared
to the halo size, while the typical haloes accrete from a wide angle in
a practically spherical pattern.
Assuming that at any given epoch the
accretion rate of dark matter, $\dot M$, is roughly proportional to the halo
mass $\Mv$ (Eq.~1 of the Letter),
while the virial densities in haloes of all masses are the same (by
definition),
the geometrical difference implies that
the densities in the filaments penetrating $\Mv\!\gg\!\mps$ haloes are higher
by a factor of a few than the typical densities in their host haloes.
The above is demonstrated in \fig{aitoff}
(as well as in Fig.~5 of Ocvirk, Pichon \& Teyssier 2008).

Assuming that the density of the gas flowing along the filaments
scales with the dark-matter density, and that the infall velocity
is comparable to the halo virial velocity,
one expects the cooling rate in the filaments feeding an $\Mv\!\gg\!\mps$
halo to be higher by a factor of a few than in the surrounding 
spherical halo.  If the compression rate in the filaments is comparable 
to that in the host halo, this implies that the thin filaments should have 
a harder time supporting a stable shock. As a result, the critical halo 
mass for shock heating in the filaments feeding it must be larger by a 
factor of a few.
This is the case for $\Mv\!\gsim\!\msh$ haloes at high redshifts.
 
A crude estimate along the lines above 
led DB06 to the conjectured upper limit for penetrating 
streams shown in \fig{coolzmf}:
\be
\mst \sim \frac{\msh}{f\mps}\msh  \quad {\rm for} \quad f\mps<\msh \ ,
\label{eq:stream}
\ee
where the characteristic width of the streams is $\prop (f\mps)^{1/3}$,
with $f$ a factor of order a few.
At low $z$, where $f\mps\!>\!\msh$, cold flows exist only for $\Mv\!<\!\msh$.
At high $z$, where $f\mps < \msh$, cold streams appear even in $\Mv\!>\!\msh$ 
haloes where shocks have heated part of the gas, as long as $\Mv\!<\!\mst$.
The critical redshift $\zc$ separating these two regimes is defined by
\be
f\mps(\zc)=\msh \ .
\ee
This crude maximum mass for cold streams is shown in \Fig{coolzmf} for an
arbitrary choice of $f=3$.
 
A preliminary analysis of the MareNostrum simulation\cite{ocvirk08}
confirms the crude prediction of \equ{stream}, when taking into account 
the lower metallicities in the simulation compared to that assumed in the 
analytic calculation. 
The streams analysed in the current Letter, in dark-matter haloes of 
$\Mv=10^{12}\msun$ at $z=2.5$,
represent an encouraging confirmation of the validity of the DB06 conjecture. 
Further analysis in progress (T.~Goerdt et al., in preparation)
indicates, for example, that at $z=2.5$, the fraction of inflow in cold 
streams drops by a factor of three at $\Mv \simeq 2\times10^{13}\msun$,
much in the spirit of the crude prediction of \fig{coolzmf}.
The permitted cold gas supply by streams in massive haloes at high redshift,
followed by a shutdown above $\msh$ at low redshifts, turn out to provide
good match to many observed galaxy properties when these features are
incorporated in semi-analytic simulations of galaxy
formation\cite{cattaneo06,croton06,bower06,cattaneo08}.
Still, the dependence of the stream properties on redshift and halo mass 
is yet to be explored in a more quantitative way.

\section{The MareNostrum simulation}
\label{simulation}

The cosmological simulation\cite{ocvirk08,prunet08}
used in the present analysis
has been performed with the Eulerian AMR
code RAMSES\cite{teyssier02} on 2,048 processors of the MareNostrum
supercomputer.
The code simulates the coupled gas and dark-matter dynamics,
using a Particle-Mesh scheme for the dark-matter component
and a second-order Godunov scheme for the gas component.
In order to describe the formation of dense star-forming disks,
the code includes metal-dependent radiative cooling,
UV heating by a standard photo-ionizing background,
star formation, supernovae feedback and metal enrichment.
The simulation box of comoving $50\hmpc$ involved $1,024^3$
dark-matter particles and $4\times \!10^{9}$ gas cells.
Using a quasi-Lagrangian refinement strategy, the spatial resolution
reaches $\sim\!1\hkpc$ in physical units at all times.
The dark-matter particle mass is $\sim \!10^{7}\msun$, 
so each of the haloes studied here consists of $\sim\!10^{5}$ 
particles within the virial radius.
Since one can reliably describe the formation of haloes down to $\sim\!100$
particles\cite{rasera06}, namely $\sim\!10^{9}\msun$, 
the $10^{12}\msun$ haloes addressed here are three 
orders of magnitude above the minimum halo mass.
This simulation allows us to capture the important properties
of gas accretion into galaxies in more than 100 haloes of $\sim\!10^{12}\msun$
at $z\!\sim\!2.5$, thus providing a large statistical sample.
A first analysis of galaxies from this simulation\cite{ocvirk08}
have confirmed the bi-modal nature of cold flows and hot media as a
function of mass and redshift\cite{db06}.

Our current analysis is based on robust features that are properly simulated,
such as the large-scale structure of the streams, the total flux in them,
and the gas clumps larger than $10^{9}\msun$. 
However, the finite resolution does not allow a fair treatment of small-scale
gas phenomena such as turbulence in the hot gas, ram-pressure stripping
of clumps, hydrodynamical instabilities at the stream boundaries,
and the formation of small clumps.
Furthermore, the current resolution does not allow a detailed study
of the disks that form at the halo centers as the disk thickness
is barely resolved.
More accurate analysis of the fine stream structure and disk buildup
should await simulations of higher resolution.

\section{Maps of entropy, flux and density}
\label{sec:maps}

\begin{figure}
\vskip 14cm
{\includegraphics{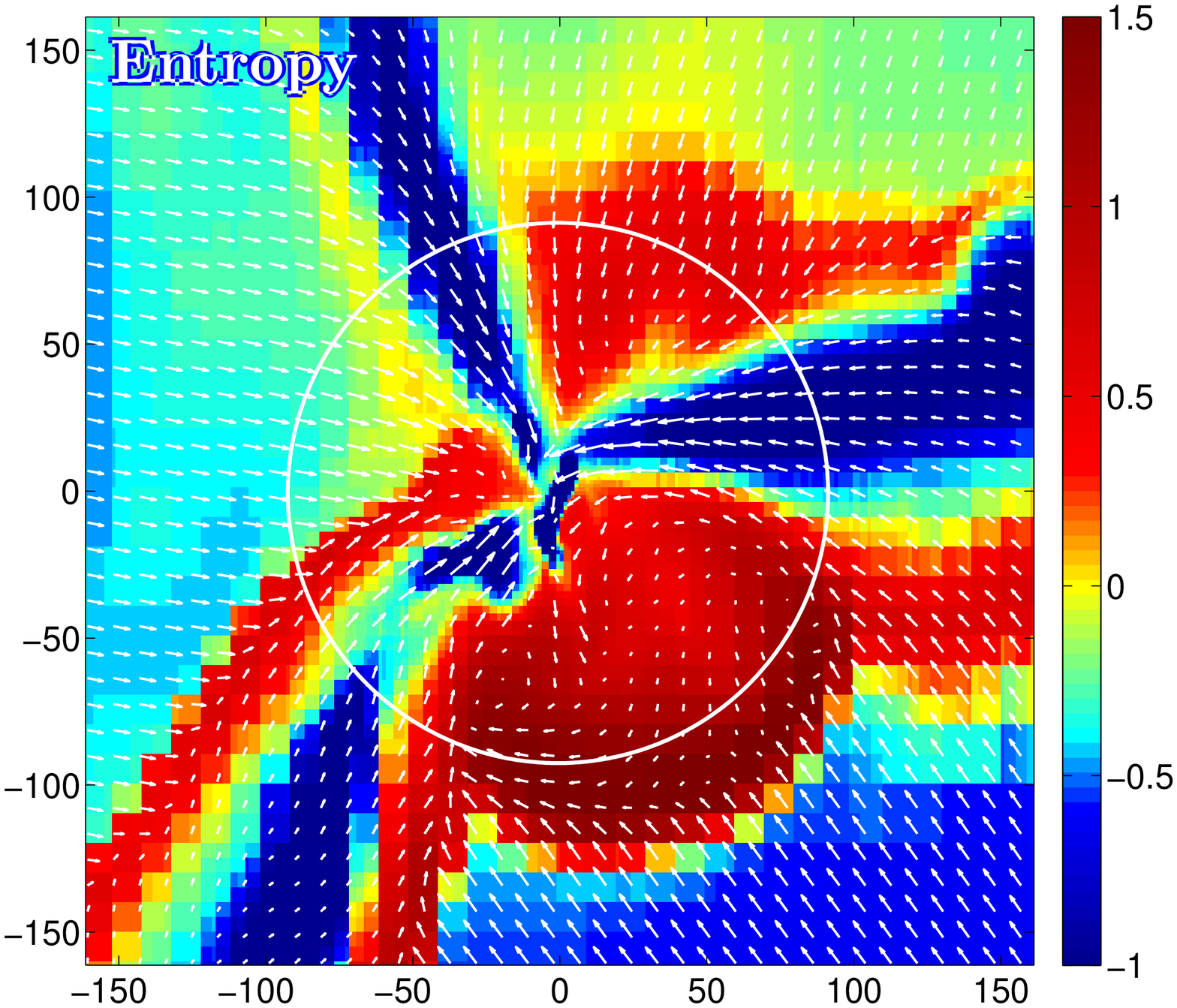}}
{\includegraphics{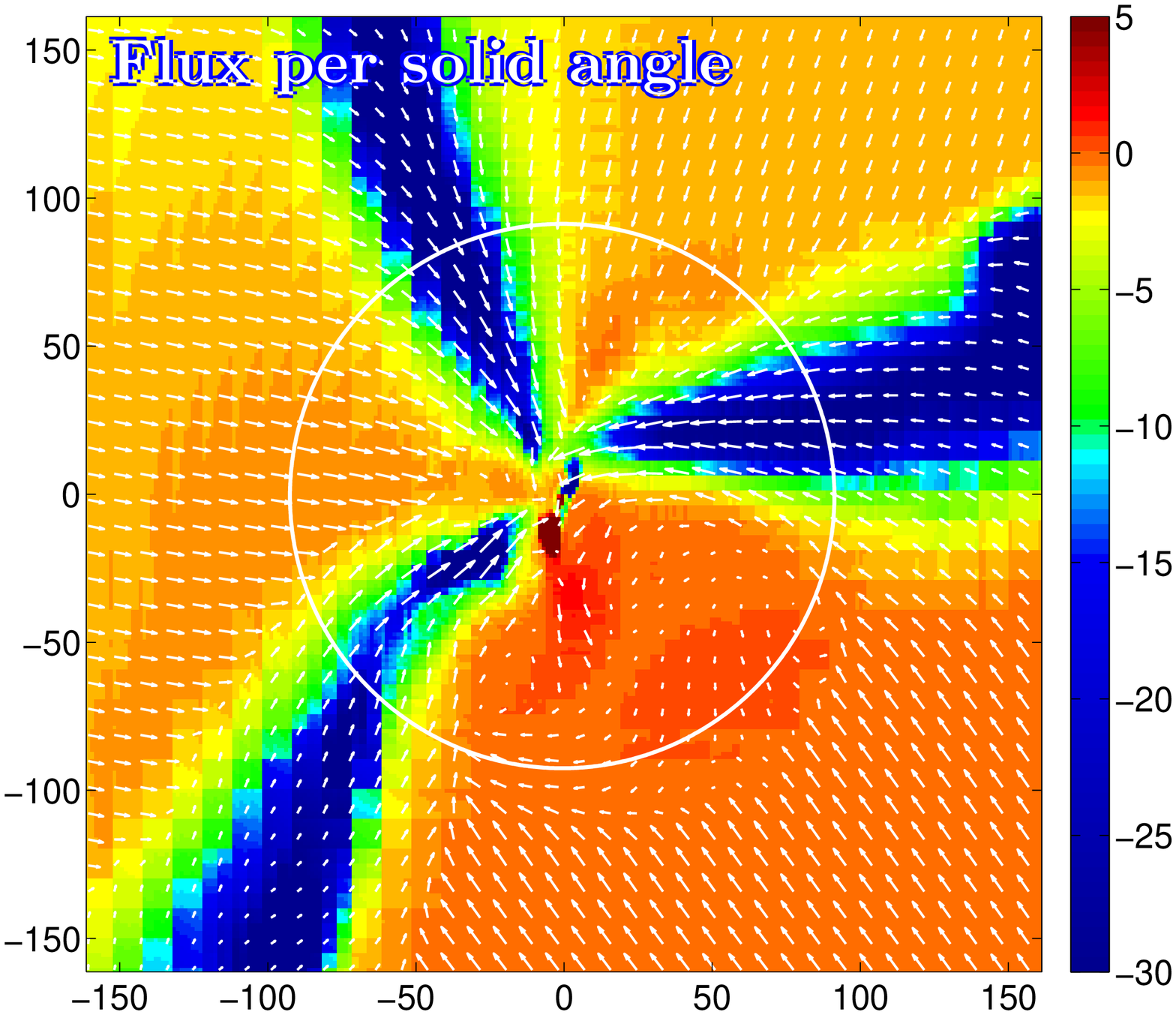}}
{\includegraphics{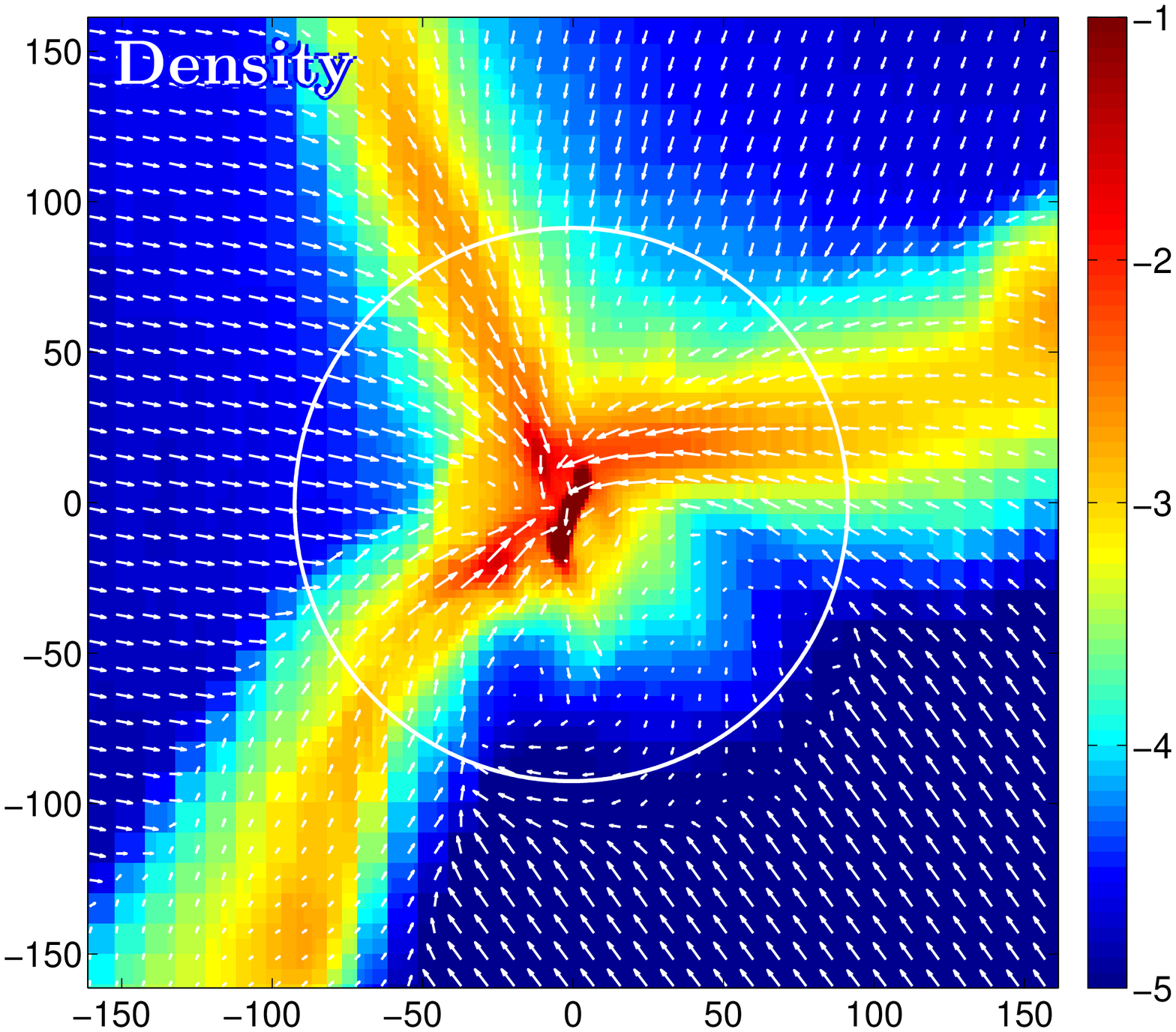}}
{\includegraphics{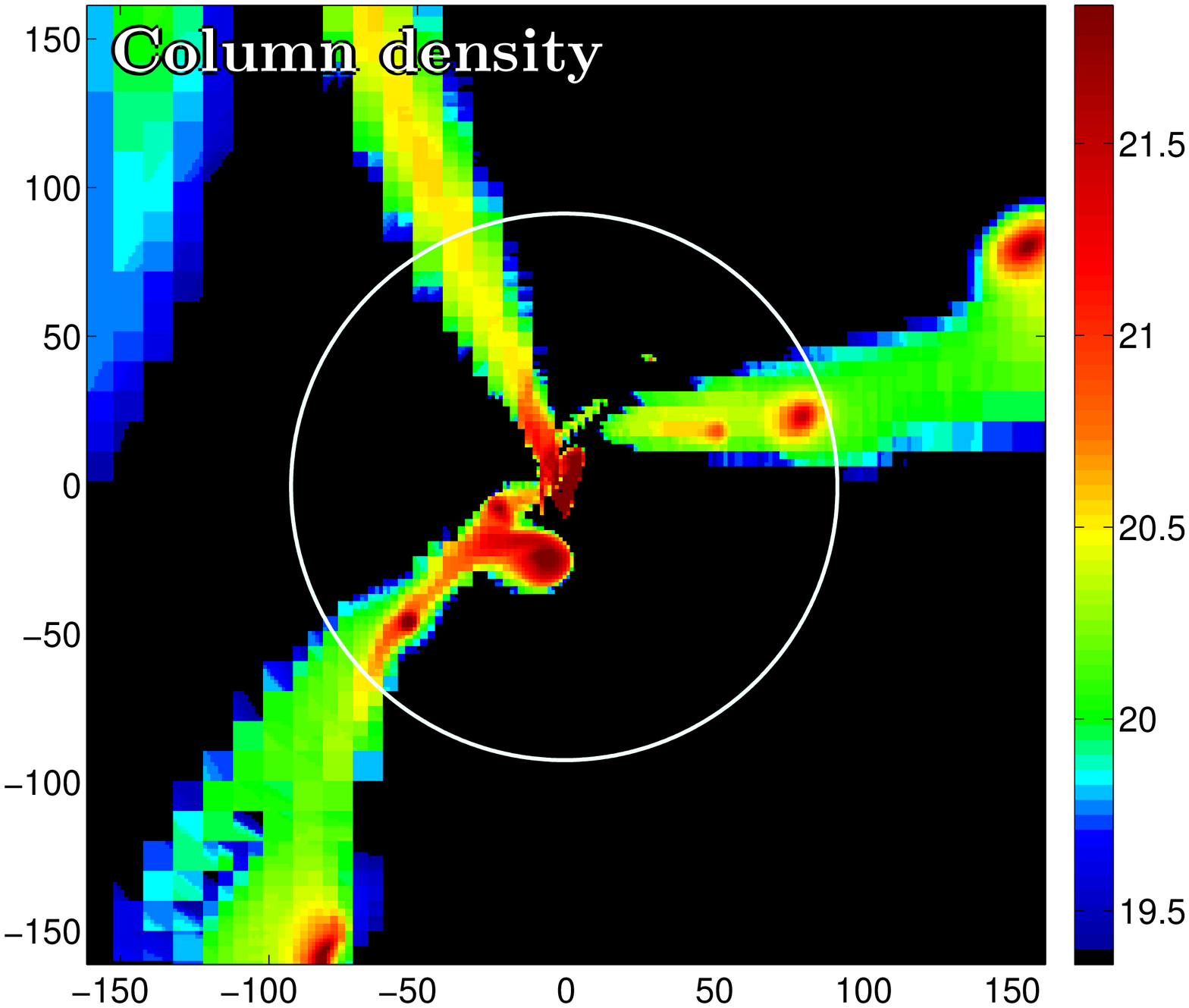}}
\vskip 0.2cm
\caption{Gas in halo 314 of the MareNostrum simulation. 
Three maps refer to a thin equatorial slice.
They show 
(a) entropy $\log K = \log (T/\rho^{2/3})$ in units of the virial quantities, 
(b) radial flux $\dot{m}=r^2 \rho\, v_r$ in $\!\sy\,{\rm rad}^{-2}$,
and (c) log density in atoms per ${\rm cm}^{-3}$. 
The fourth panel shows log column density through the 3D box of side $320\kpc$,
in ${\rm cm}^{-2}$, 
considering only the cells where the radial flux inward is at least twice 
as high as the average over a shell based on Eq.~1 of the Letter.
The circle marks the virial radius.
}
\label{fig:314}
\end{figure}

\begin{figure}
\vskip 14cm
{\includegraphics{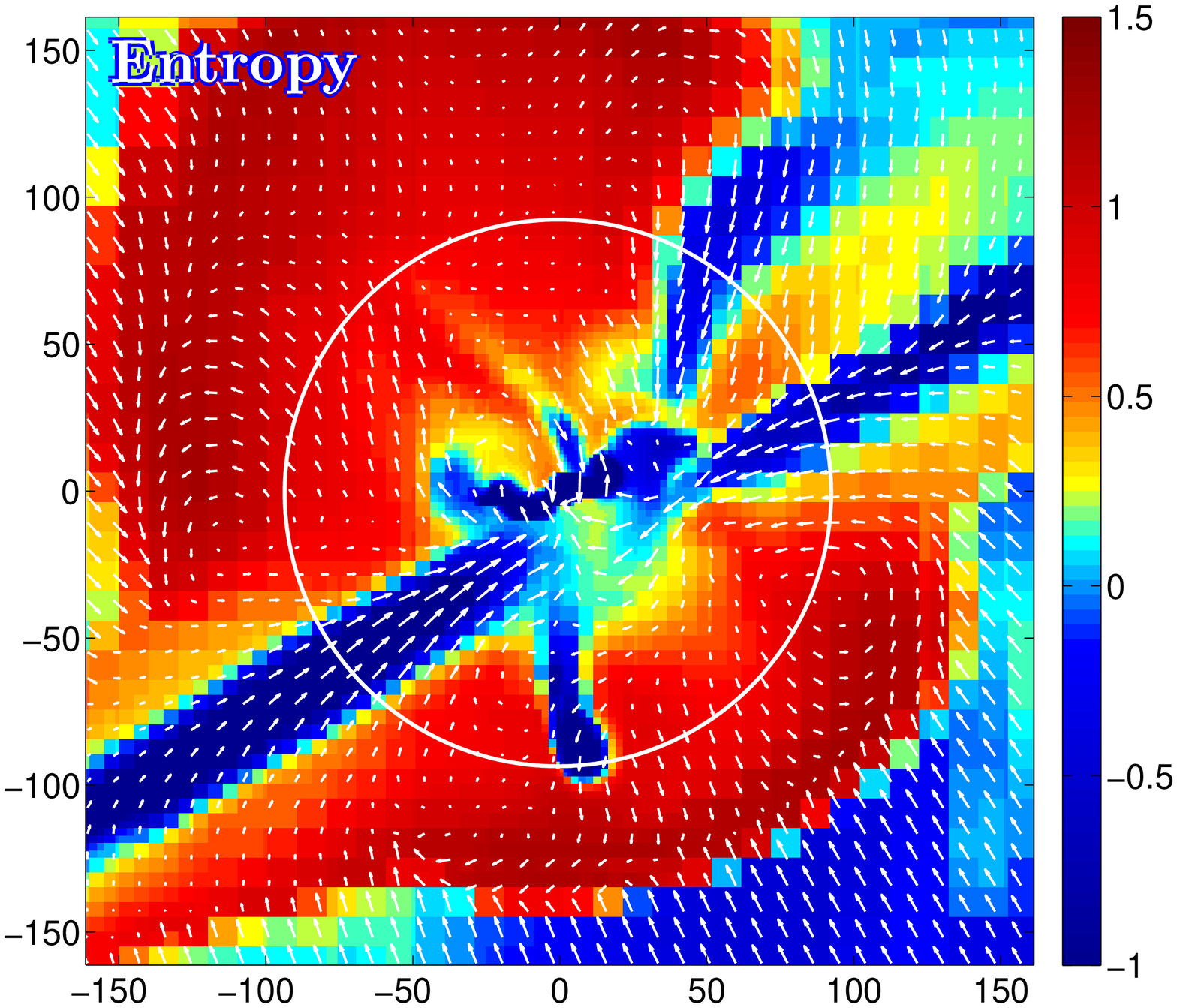}}
{\includegraphics{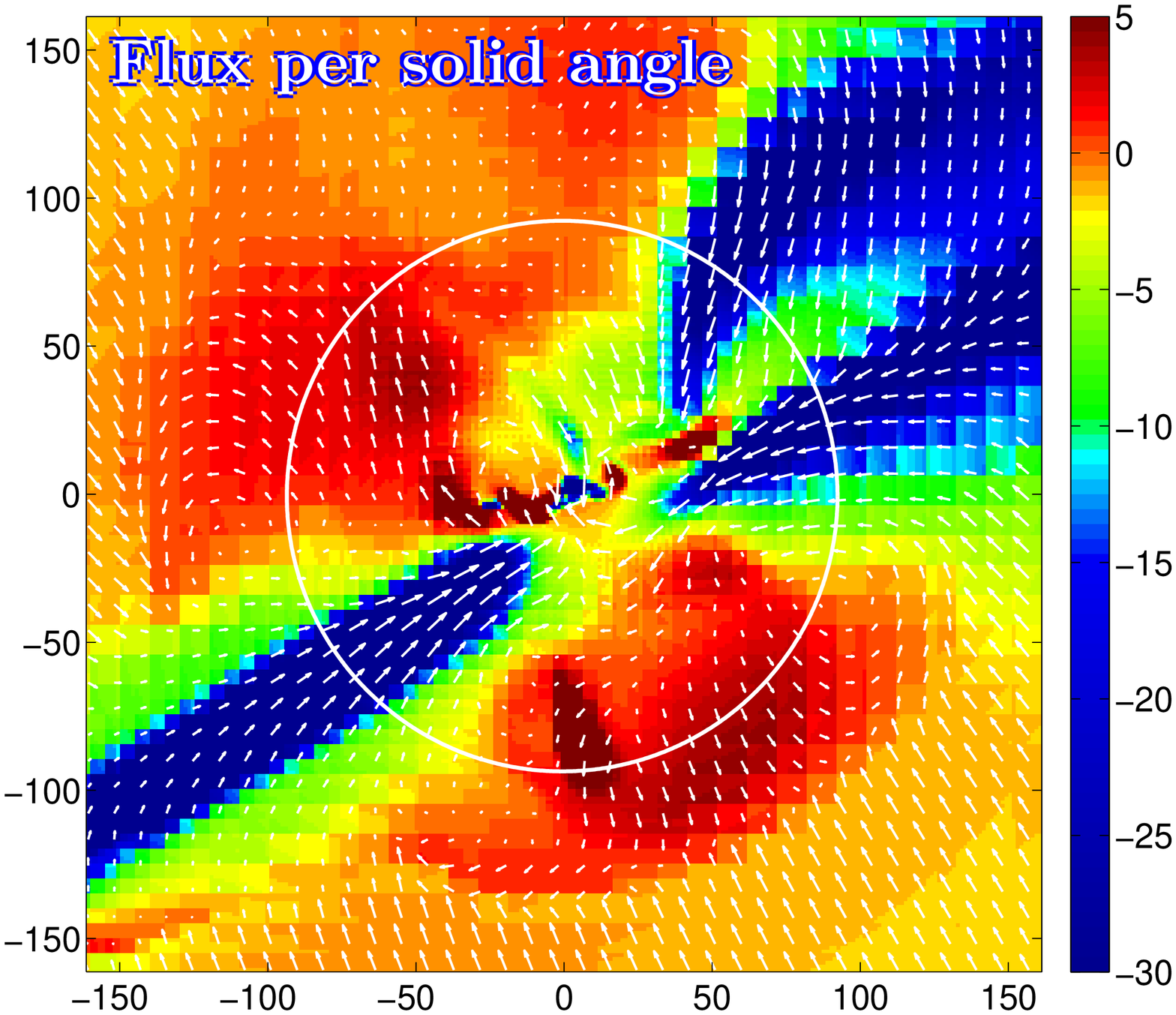}}
{\includegraphics{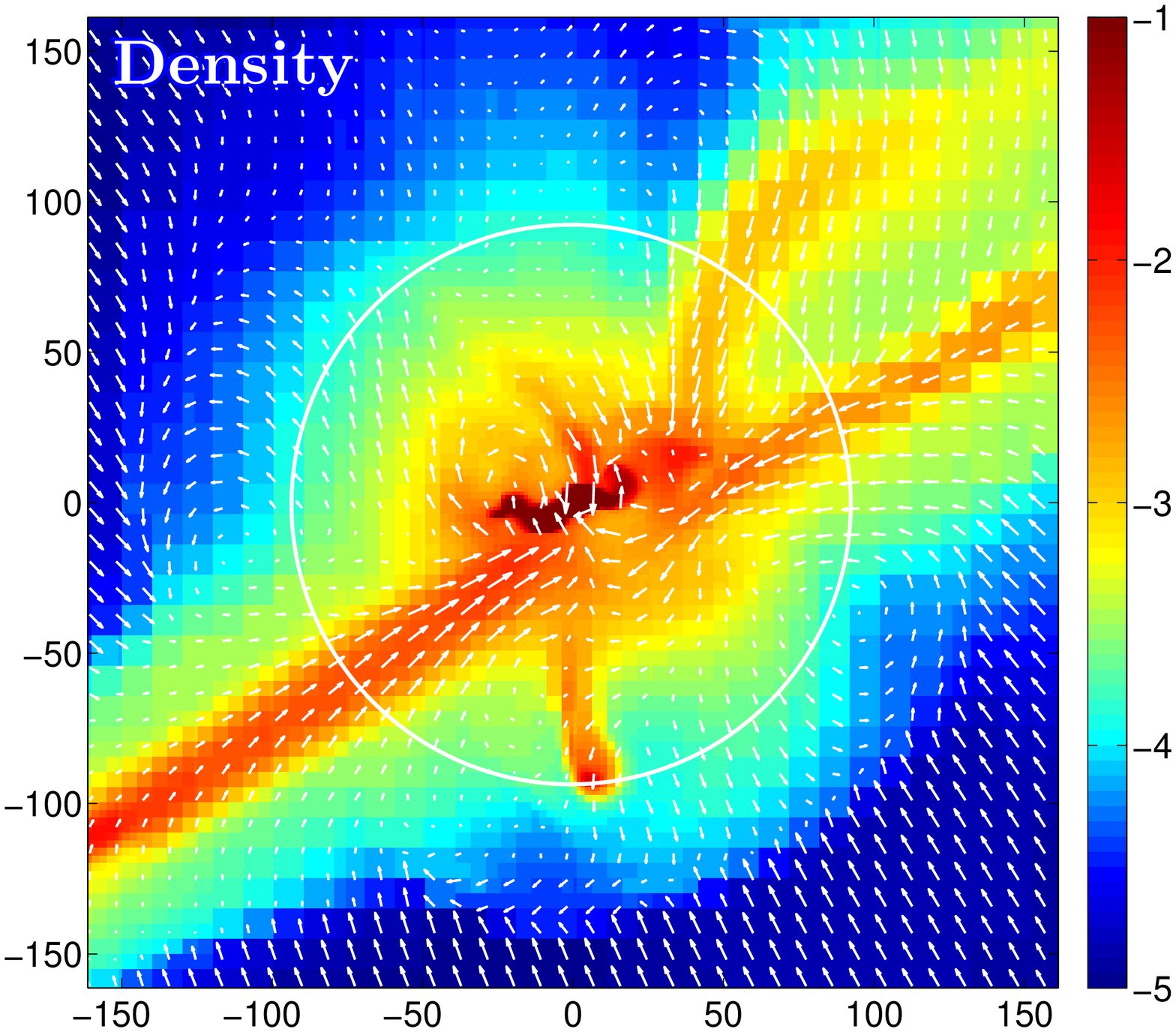}}
{\includegraphics{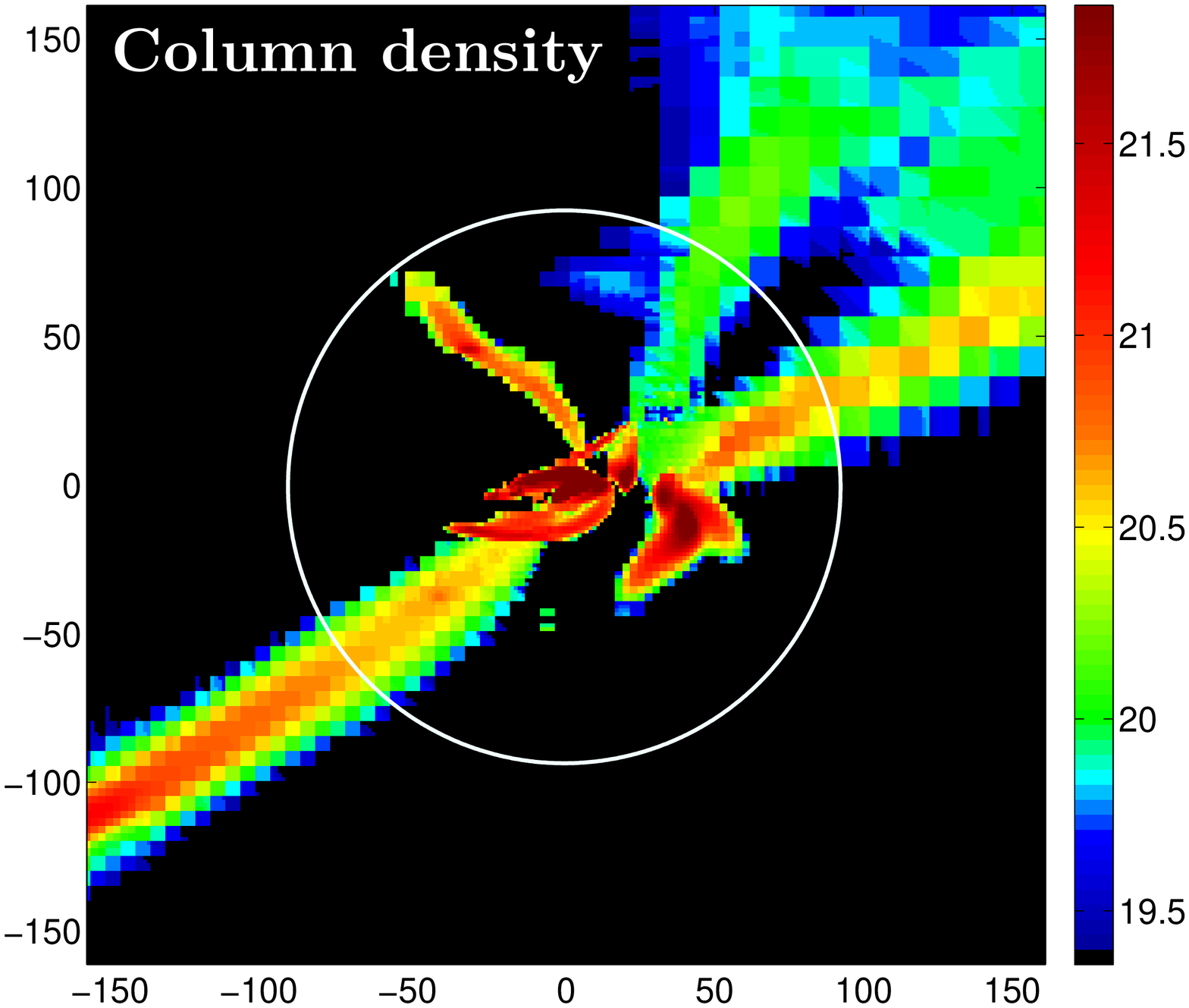}}
\vskip 0.2cm
\caption{Gas in halo 303 of the MareNostrum simulation.
See \fig{314}.
}
\label{fig:303}
\end{figure}

\begin{figure}
\vskip 14cm
{\includegraphics{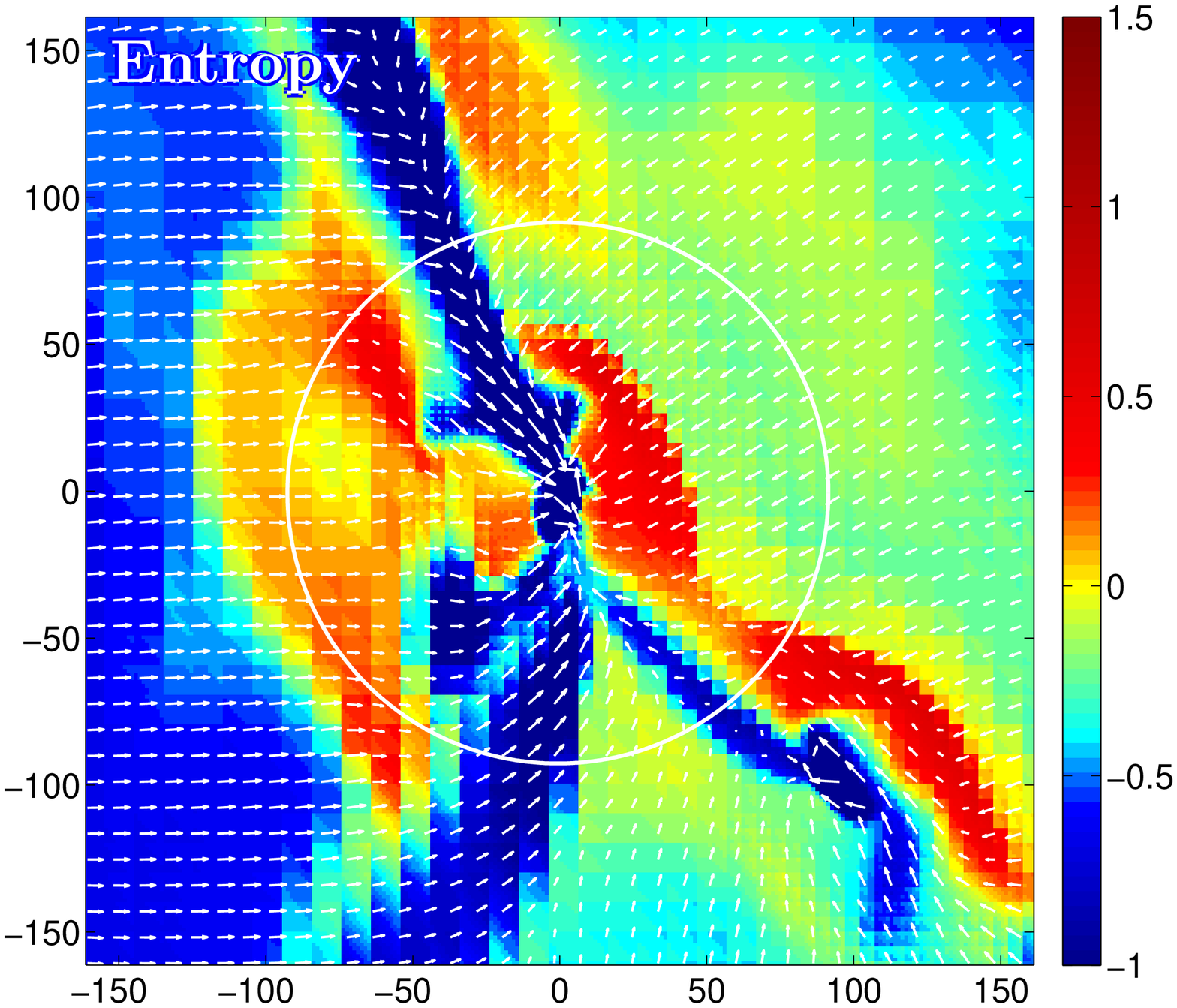}}
{\includegraphics{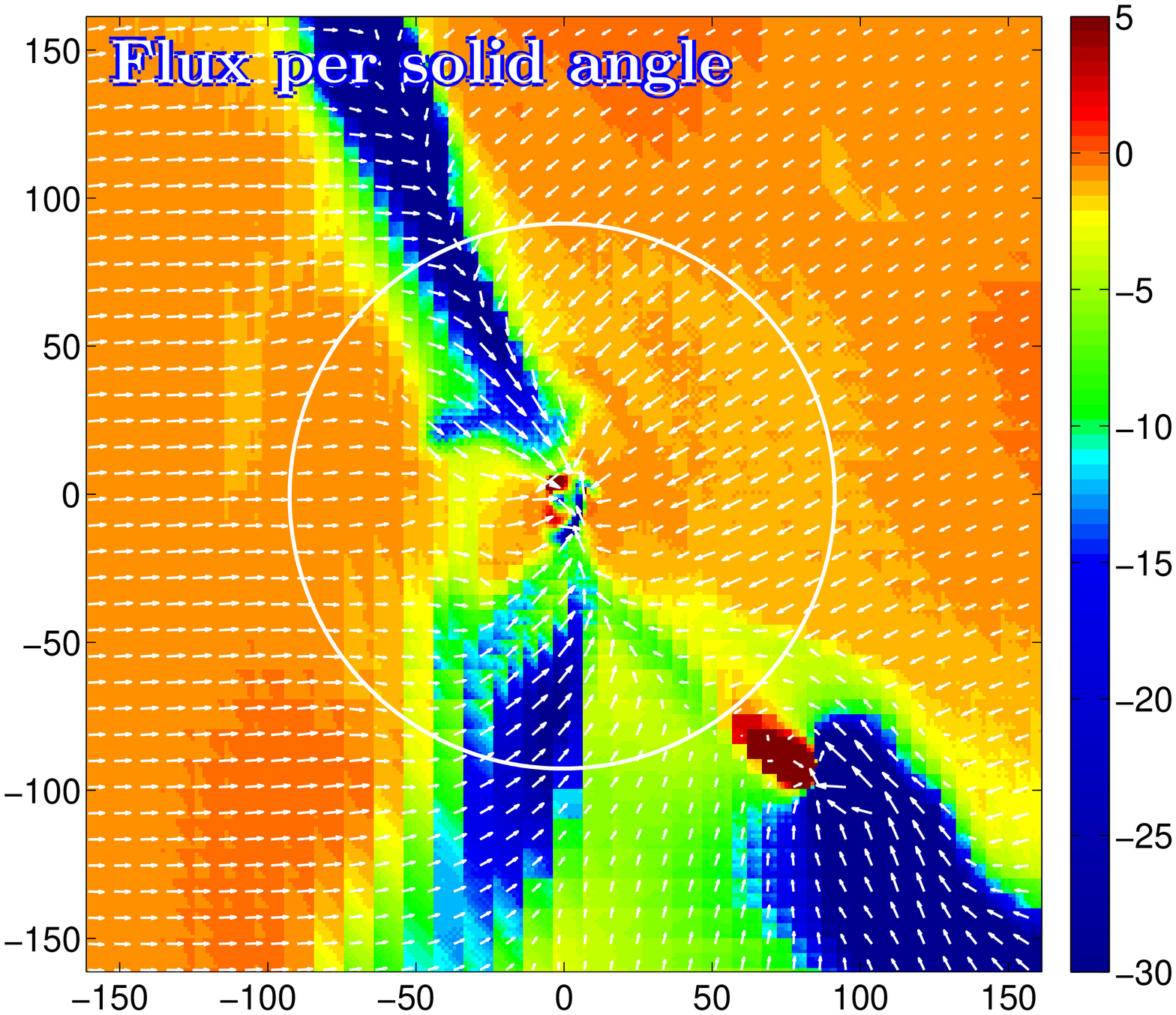}}
{\includegraphics{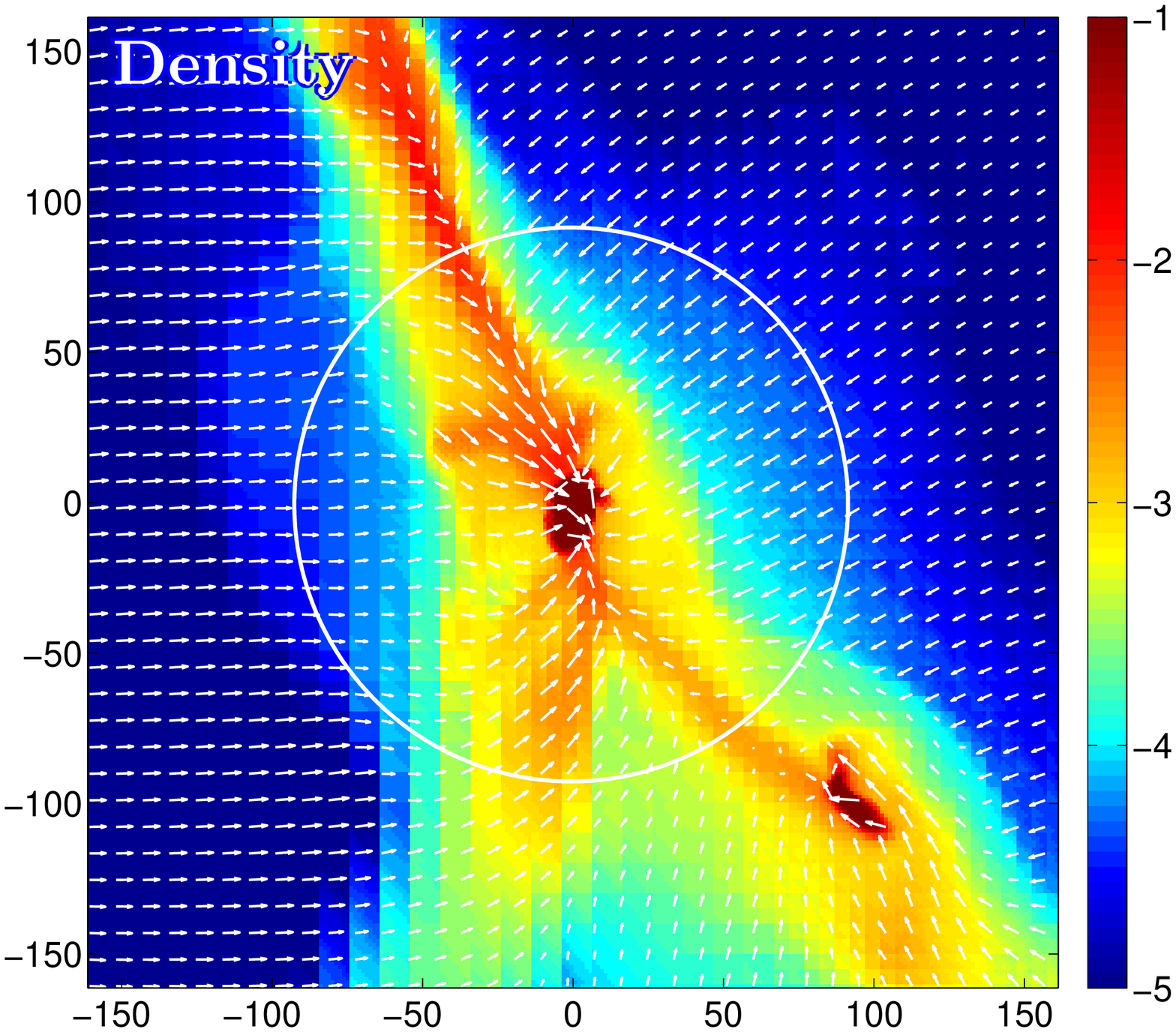}}
{\includegraphics{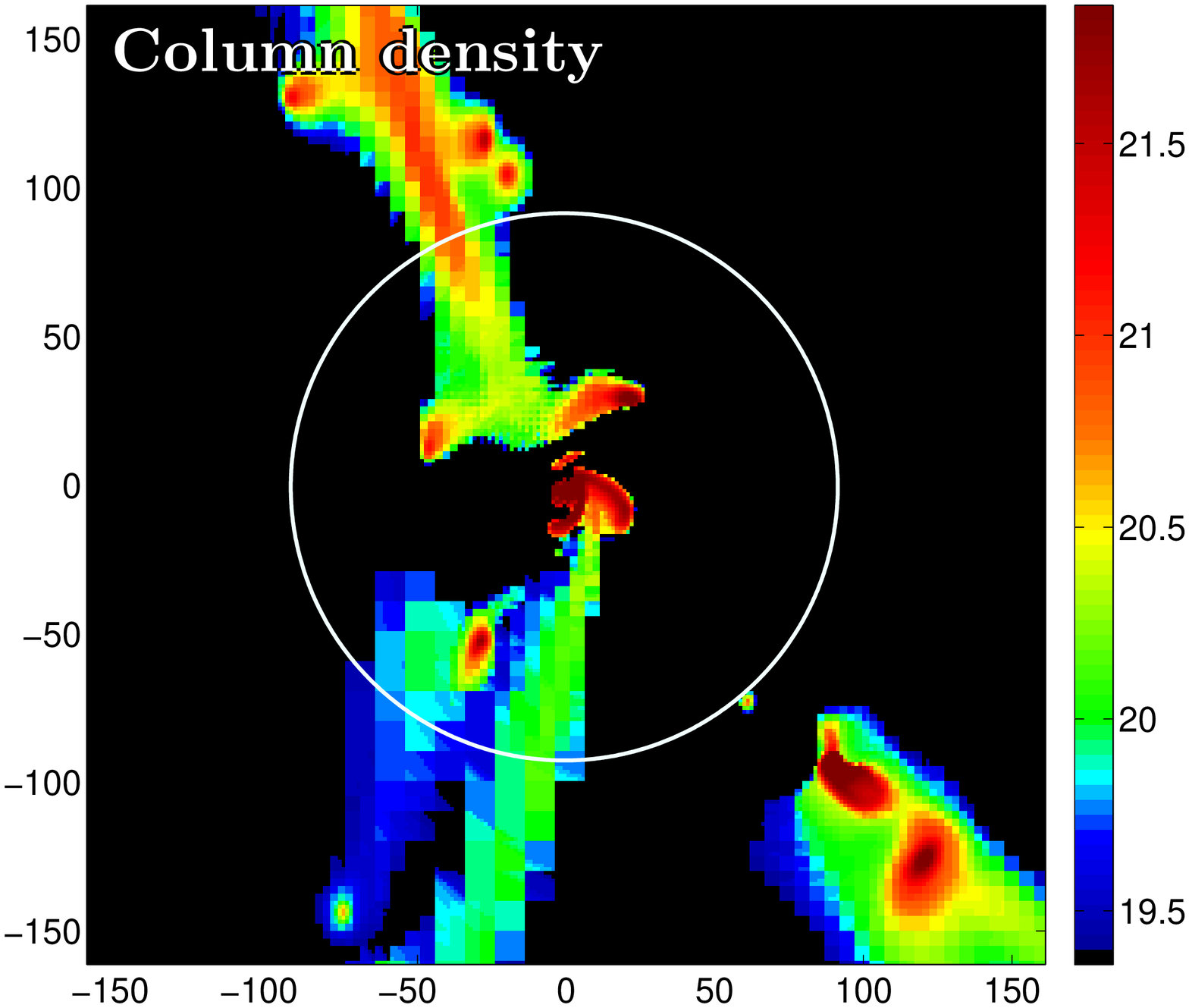}}
\vskip 0.2cm
\caption{Gas in halo 311 of the MareNostrum simulation.
See \fig{314}.
}
\label{fig:311}
\end{figure}

\Figs{314} to \ref{fig:311} extend the visual information provided
by Figs.~1 and 2 of the Letter. They display different gas properties 
that highlight the structure and kinematics of the cold streams in three
simulated galaxies of $\Mv\!=\!10^{12}\msun$ at $z\!=\!2.5$.

The entropy maps show $\log (T/\rho^{2/3})$ 
where the temperature and gas density
are in units of the virial temperature and mean density within the halo virial
radius $\Rv$. They exhibit the virial shock, covering most of the area of the
virial sphere and sometimes extending beyond $2\Rv$. The narrow streams
are of much lower entropy, by more than three orders of magnitude, 
comparable to the low entropy in the central disk they lead to. 
The boundaries between the streams and the hot medium within the virial radius
are sharp and well defined.  
We also note that semi-cylindrical shocks sometimes partly surround the 
elongated streams long before they enter the halo virial radius. 

The arrows mark the velocity field projected on the slice plane,
and the flux colour maps show the flow rate per solid angle,
$\dot{m}=r^2\rho\,v_r$. 
The flux inward is almost exclusively channeled through the narrow streams,
typically involving 95\% of the total inflow rate.
This flux is several times the average over a sphere, $\dot{m}_{\rm
vir}\!\simeq\!8\sy {\rm rad}^{-2}$. 
The opening angle of a typical stream at $\Rv$ is $20-30^\circ$,
so the streams cover a total area of $\sim\!0.4\,{\rm rad}^2$, namely
a few percent of the sphere.
The velocity field in the hot medium is turbulent and sometimes showing vast
outflows. The inward flux over most of the sphere area is negligible,
both inside and outside the virial radius or the virial shock. 
The streaming velocities are supersonic, with a Mach number of order a few.

Although the streams tend to be rather radial when viewed on scales
comparable to the halo virial radius, some of them flow in with impact 
parameters on the order of $10\kpc$, comparable to the disk sizes. 
The steady high flux along a line of a rather fixed orientation
with a non-negligible impact parameter is the source of  
angular momentum required for the buildup of an extended rotating disk
(A. Zinger et al., in preparation).

The gas density maps emphasize the narrowness of the streams, and reveal that
they are typically denser than the surrounding medium by more than an order 
of magnitude. This confirms the prediction described in \se{origin},
and explains why a virial shock is avoided along the streams, 
allowing them to penetrate cold and unperturbed into the inner halo.

The column-density maps of the in-flowing material 
are obtained by summing up the densities in grid cells along
each line of sight inside the box of side $320\kpc$. 
The cells that enter this sum are only those where the inward flux per 
solid angle is at least twice the average over a sphere based on Eq.~1 
of the Letter. 
These maps highlight the 
three-dimensional configuration of radial streams, and the clumps
along some of them.
Such column density maps will serve us in producing
detailed predictions to guide observations in search of  
the cold streams in high-redshift galaxies, in absorption and in emission.

\begin{figure}
\vskip 14.5cm
{\includegraphics{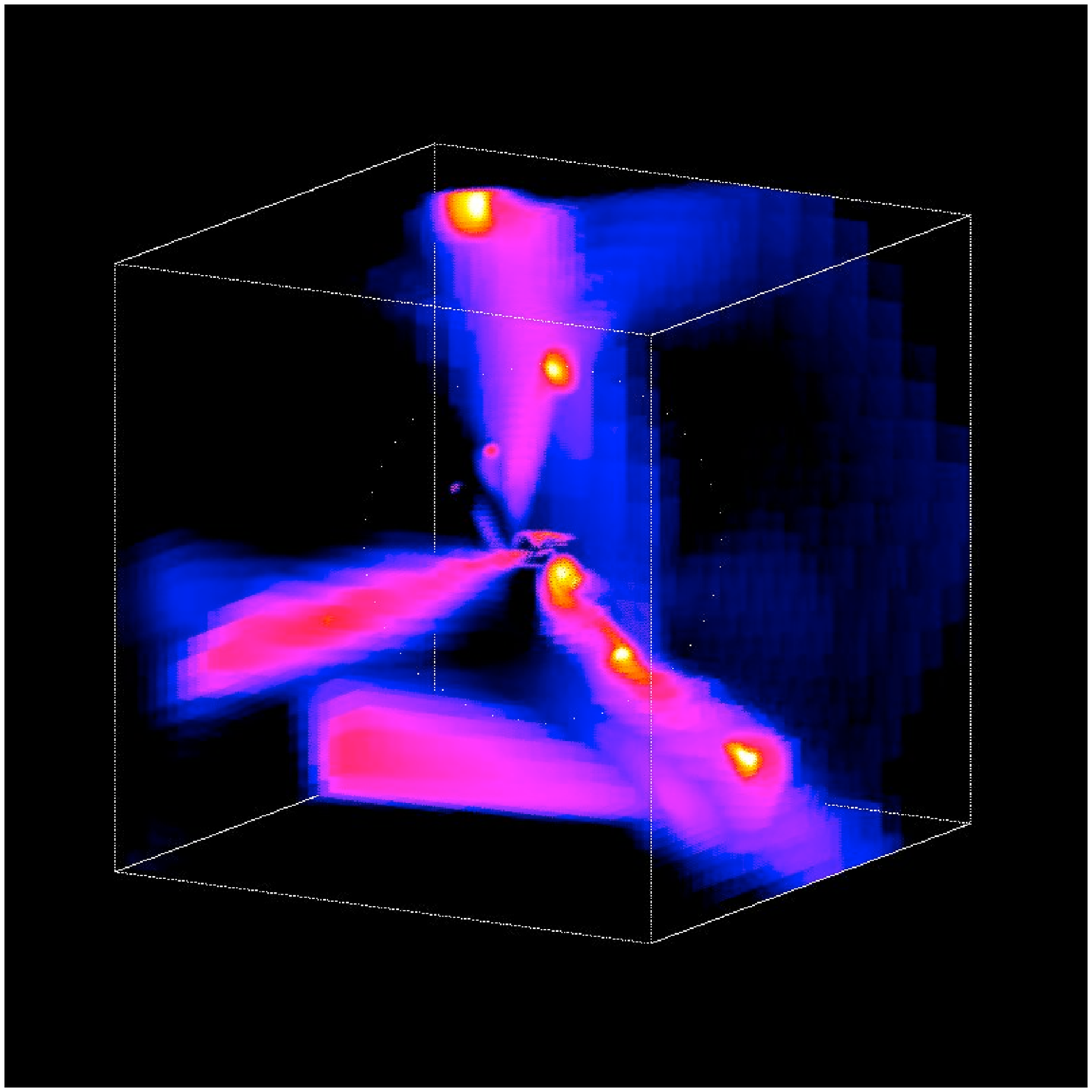}}
{\includegraphics{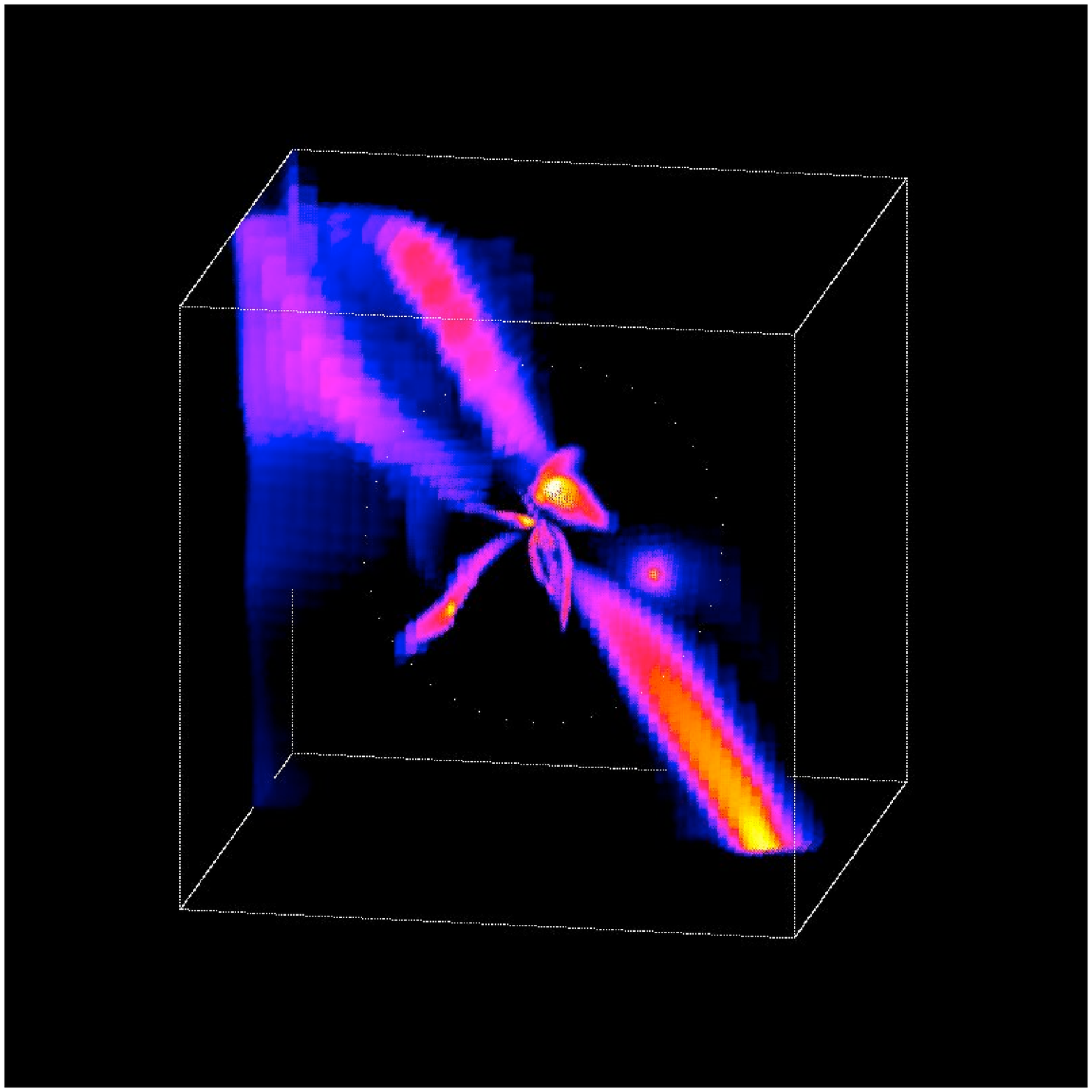}}
{\includegraphics{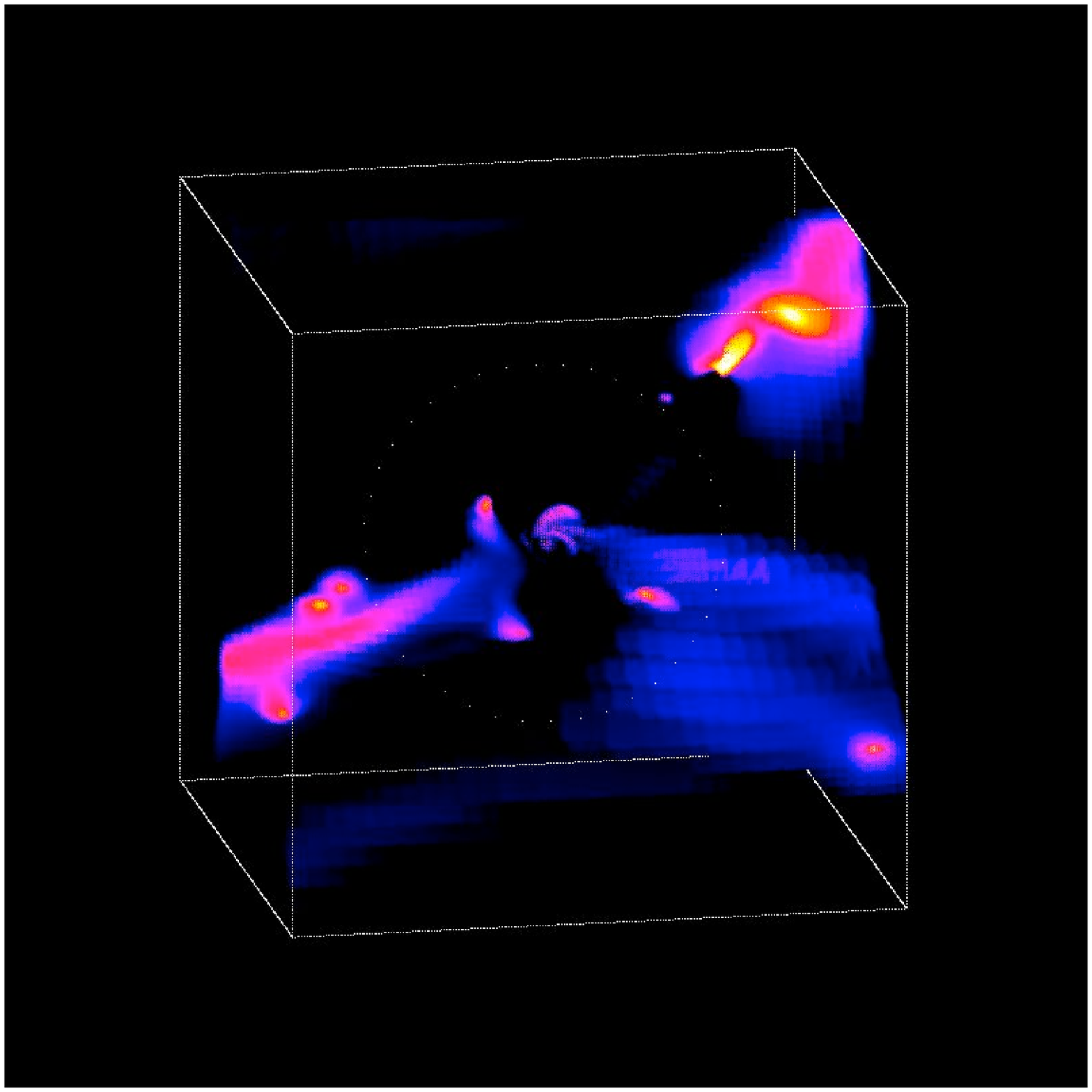}}
{\includegraphics{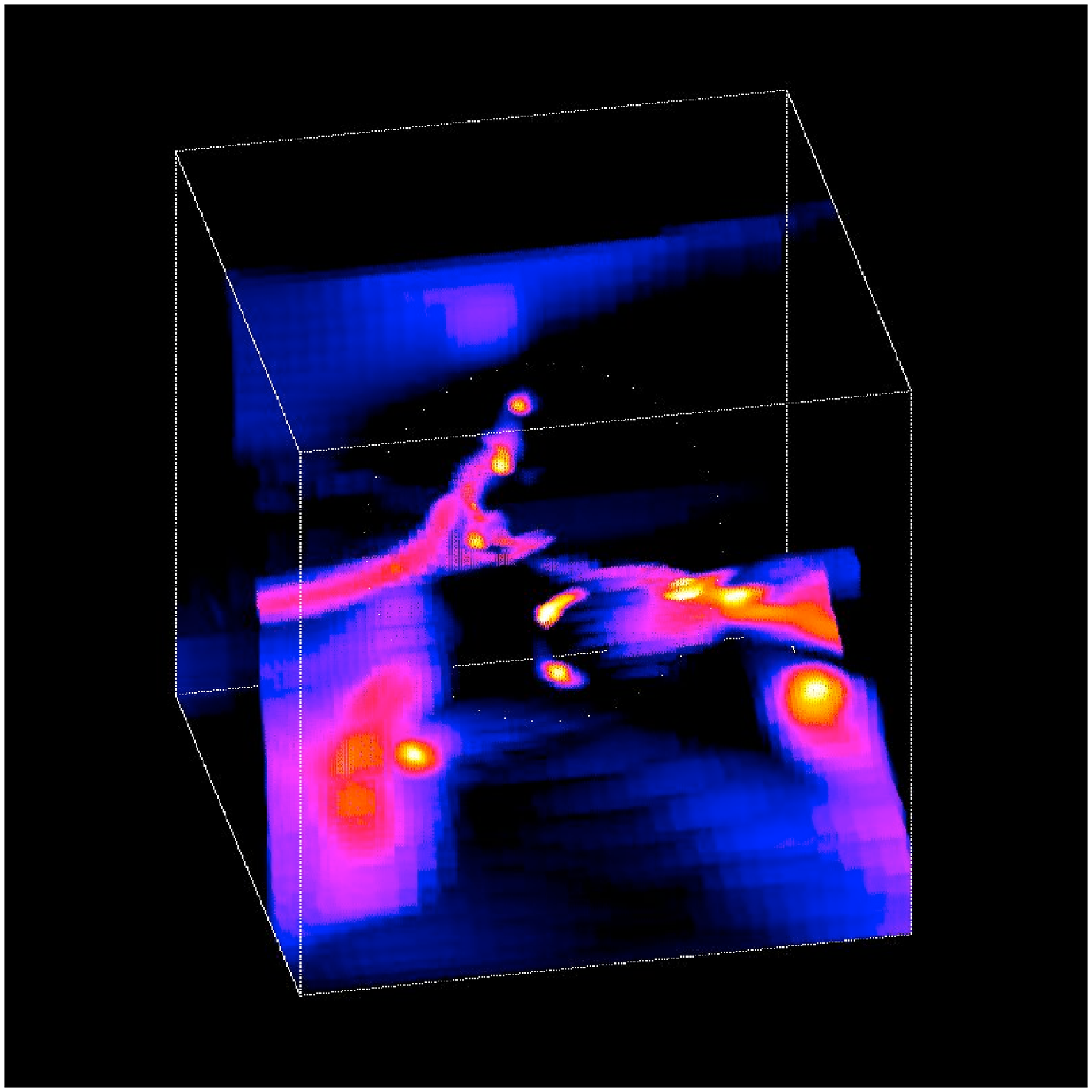}}
\vskip 0.2cm
\caption{Inward flux in the three-dimensional boxes of side
$320\kpc$ centered on galaxies 314, 303, 311 and 310 from the MareNostrum
simulations.  The colours refer to inflow rate per solid angle of point-like
tracers at the centers of cubic-grid cells. The dotted circle marks the
virial radius.  All haloes show high-flux
streams, some smooth and some with embedded clumps. 
Galaxy 310 (bottom right) is undergoing multiple minor mergers 
due to the particularly clumpy streams.
}
\label{fig:3D}
\end{figure}

\Fig{3D} displays three-dimensional
TIPSY\footnote{http://www-hpcc.astro.washington.edu/tools/tipsy/tipsy.html}
pictures of the radial influx $\dot{m}$, similar to Fig.~2 of the Letter.
It shows the overall structure of the in-flowing streams in 3D perspective
for four simulated haloes. The pictures reveal that the typical configuration 
is of three major narrow streams. Some of the streams are straight lines,
and others are curved. Some of the streams are of rather fixed width from
well outside $\Rv$, and others display a conical shape, 
starting broad at large radii and getting narrower as they penetrate into
the halo. The gas streams show dense clumps, with about 
one third of 
the stream mass in 
clumps of mass ratio $\mu\!>\!0.1$, namely mass above $\sim\!10^{10}\msun$. 
The rest is in smaller clumps, some clearly hidden below the
resolution limit. Since these mini-minor clumps are not expected to cause
significant damage to the central disk\cite{cox08}, we can refer to them 
in this respect as ``smooth" flows.  
It is not clear at this point to what extent the smooth component
is truly smooth or built by mini-minor clumps,
and whether the perfect smoothness has a physical origin or is merely 
a numerical artifact, but this distinction does not make
a qualitative difference to our present diskussion.

\vfill\eject
\section{Accretion profiles and probability distribution}

\begin{figure}[h]
\vskip 12.0cm
{\includegraphics{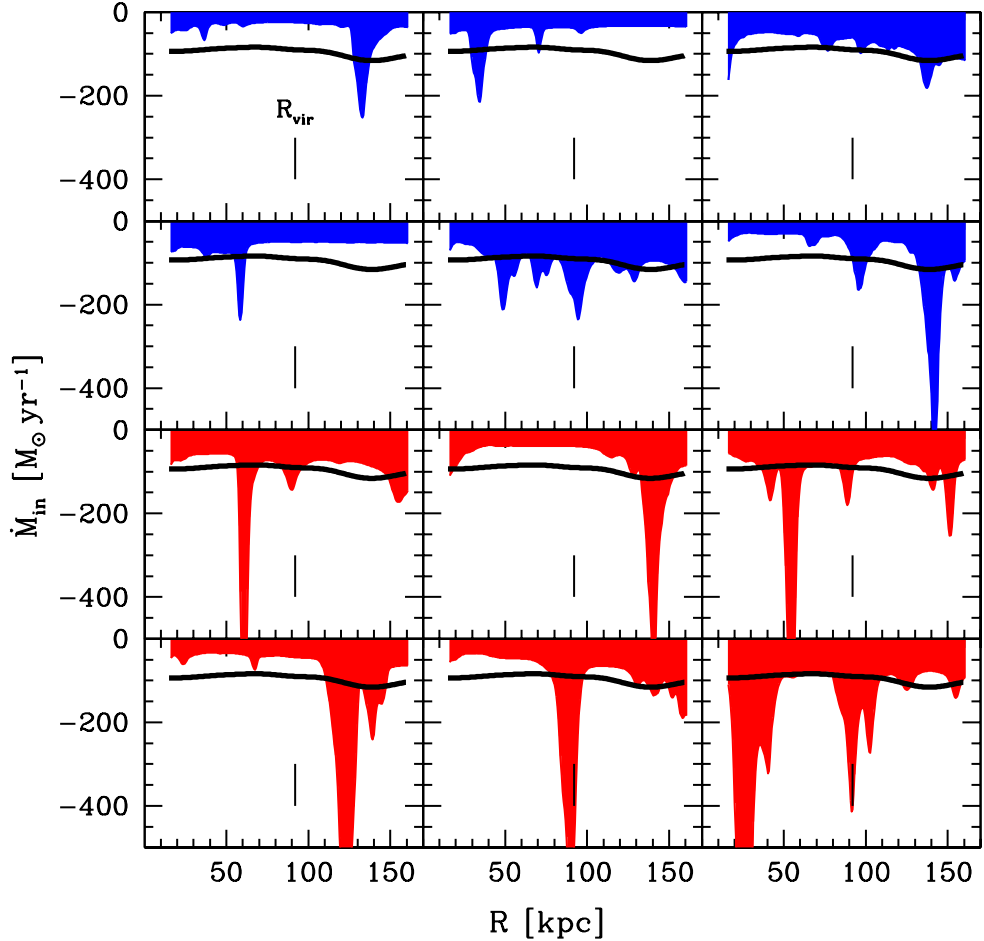}}
\vskip 0.2cm
\caption{Profiles of total gas inflow rate through spherical shells
as in Fig.~3 of the Letter. Shown here are twelve galaxies of
$\Mv\simeq 10^{12}\msun$ at $z=2.5$, randomly chosen from the simulation.
The lower six panels show clumps that correspond to mergers of mass ratio
$\mu > 0.1$, while the upper six are fed by smoother flows with only
mini-minor mergers of $\mu < 0.1$.
}
\label{fig:profiles}
\end{figure}

\Fig{profiles} is an extension of Fig.~3 of the Letter, 
presenting the influx profiles of twelve galaxies, 
all with $\Mv\!\simeq\!10^{12}\msun$ at $z\!=\!2.5$,
chosen at random from the MareNostrum simulation.
The profiles extend from $r\!=\!15\kpc$, the disk vicinity,  
to $r\!=\!160\kpc$, almost twice the virial radius of $\Rv\!\simeq\!90\kpc$.



\begin{figure}
\vskip 8.5cm
{\includegraphics{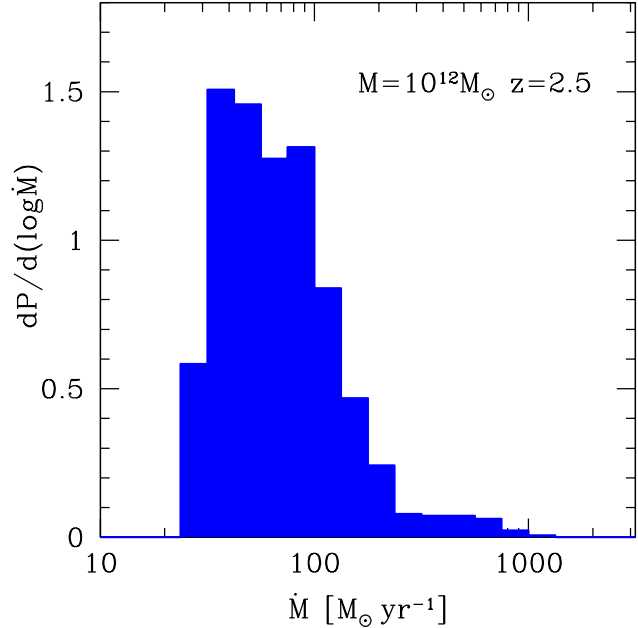}}
\vskip 0.2cm
\caption{The conditional probability distribution $P(\dot{M}|\Mv)$
for the fiducial case $\Mv\!=\!M_0\!=\!10^{12}\msun$ at $z\!=\!2.5$.
}
\label{fig:distribution}
\end{figure}

In order to evaluate the conditional probability $P(\dot{M}|\Mv)$
that enters Eq.~2 of the Letter, we first measure $P_0(\dot{M}|M_0)$ 
from a fair sample of MareNostrum haloes of our fiducial case
$M_0\!=\!10^{12}\msun$ at $z_0\!=\!2.5$.
This probability distribution, shown in \fig{distribution}, is derived by 
sampling the $\dot{M}(r)$ profiles shown in \fig{profiles} uniformly in $r$,
using the fact that the inflow velocity along the streams is roughly constant.
The tail of the distribution shown in \fig{distribution},
at $\dot{M}\!>\!200\sy$,
is dominated by $\mu\!>\!0.1$ mergers,
while the main body of the distribution is mostly due to smoother streams.
Recall that the average is about $100\sy$.

We then generalize $P_0(\dot{M}|M_0)$ to other masses $\Mv$
using the scaling from \equ{mdot_vir_2}, $\dot{M} \propto \Mv^{1.15}$, namely
\be
P(\dot{M}|\Mv) = P_0[\dot{M} ({M_0}/{\Mv})^{1.15} |M_0] \ .
\ee
At $z\!\sim\!2.5$, this scaling of $\dot{M}$ is good to within a factor of two
for $\Mv \leq 10^{13}\msun$, beyond which it becomes a more severe
overestimate (Goerdt et al., in preparation).
The results for other redshifts ($z\!>\!2$) are obtained
using the scaling from \equ{mdot_vir_2}, $\dot{M}\propto (1+z)^{2.25}$.

Half the galaxies shown in \fig{profiles}
turn out to show clumps leading to mergers of $\mu\!>\!0.1$,
and the rest show only smaller clumps in smoother flows.
One can read from the relative width of the clumps in \fig{profiles} 
that the duty cycle for $\mu\!>\!0.1$ clumps
in each individual galaxy is less than $0.1$. 
By comparing the areas above the individual profiles
with the average for galaxies of that  
mass and redshift, one can see that on average only about 
one third of 
the stream mass is in clumps.

A similar estimate is obtained from EPS,
by reading from Fig.~6 of Neistein \& Dekel\cite{neistein08b}
the rate $dN/d\omega$ of $>\!\mu$ mergers into a halo $\Mv$.
The typical starburst duration is $\Delta t\!\simeq\! 0.1 \Rv/\Vv$
($\sim\!50 \Myr$ at $z\!=\!2.5$),
based on merger simulations\cite{cox08} or
the typical peak width in the
$\dot{M}(r)$ profiles (\fig{profiles}), given streaming at the virial velocity
$\Vv=(G\Mv/\Rv)^{1/2}\!\sim\!220\kms$.
This leads to $\eta\!=\!(dN/d\omega) \Delta t \!\simeq\! 0.09$ for
$M\!=\!2\!\times\!10^{12}\msun$ at $z\!=\!2.2$.

\section{The abundance as a function of mass and redshift}

\begin{figure}
\vskip 8.0cm
{\includegraphics{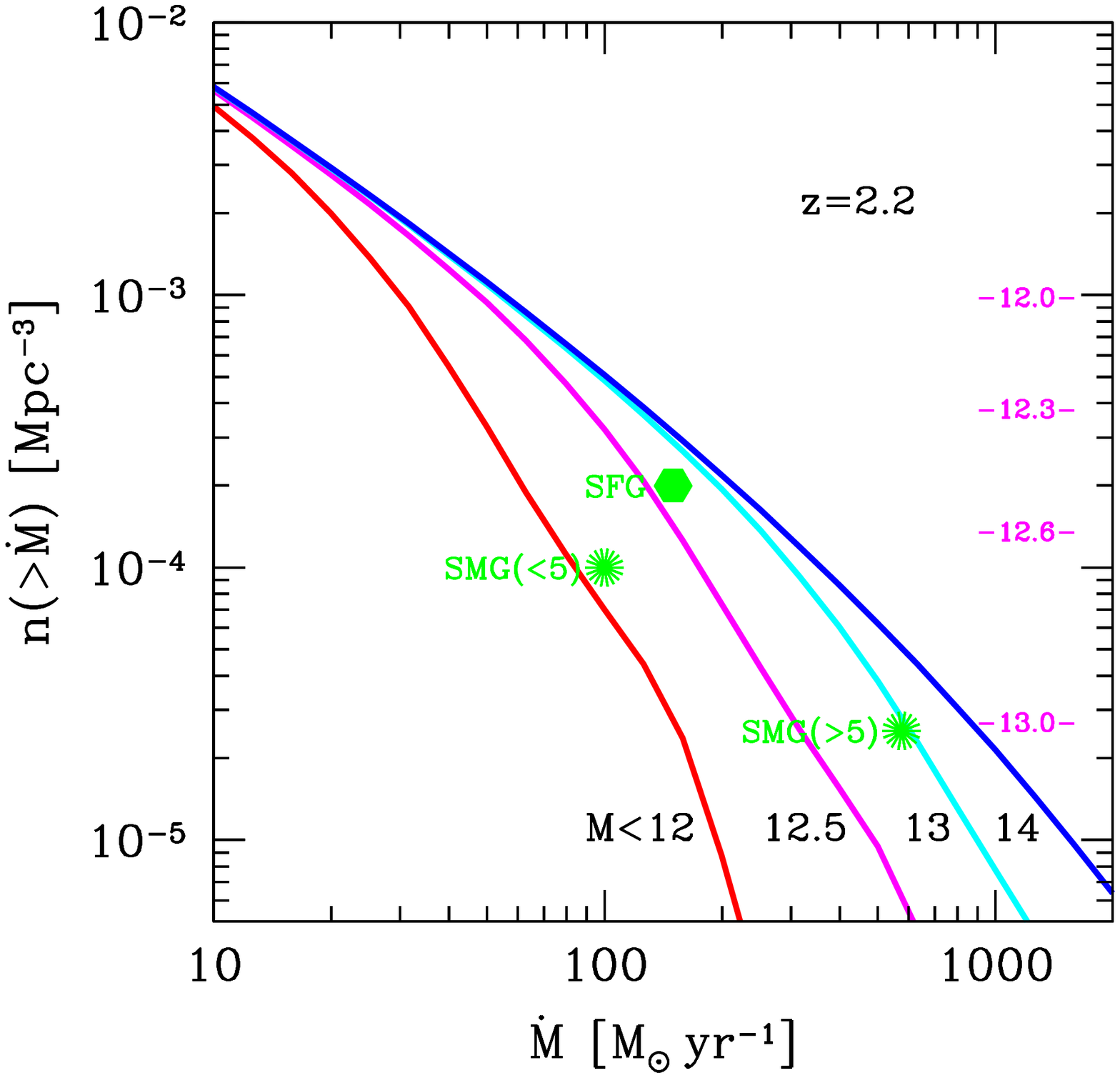}}
{\includegraphics{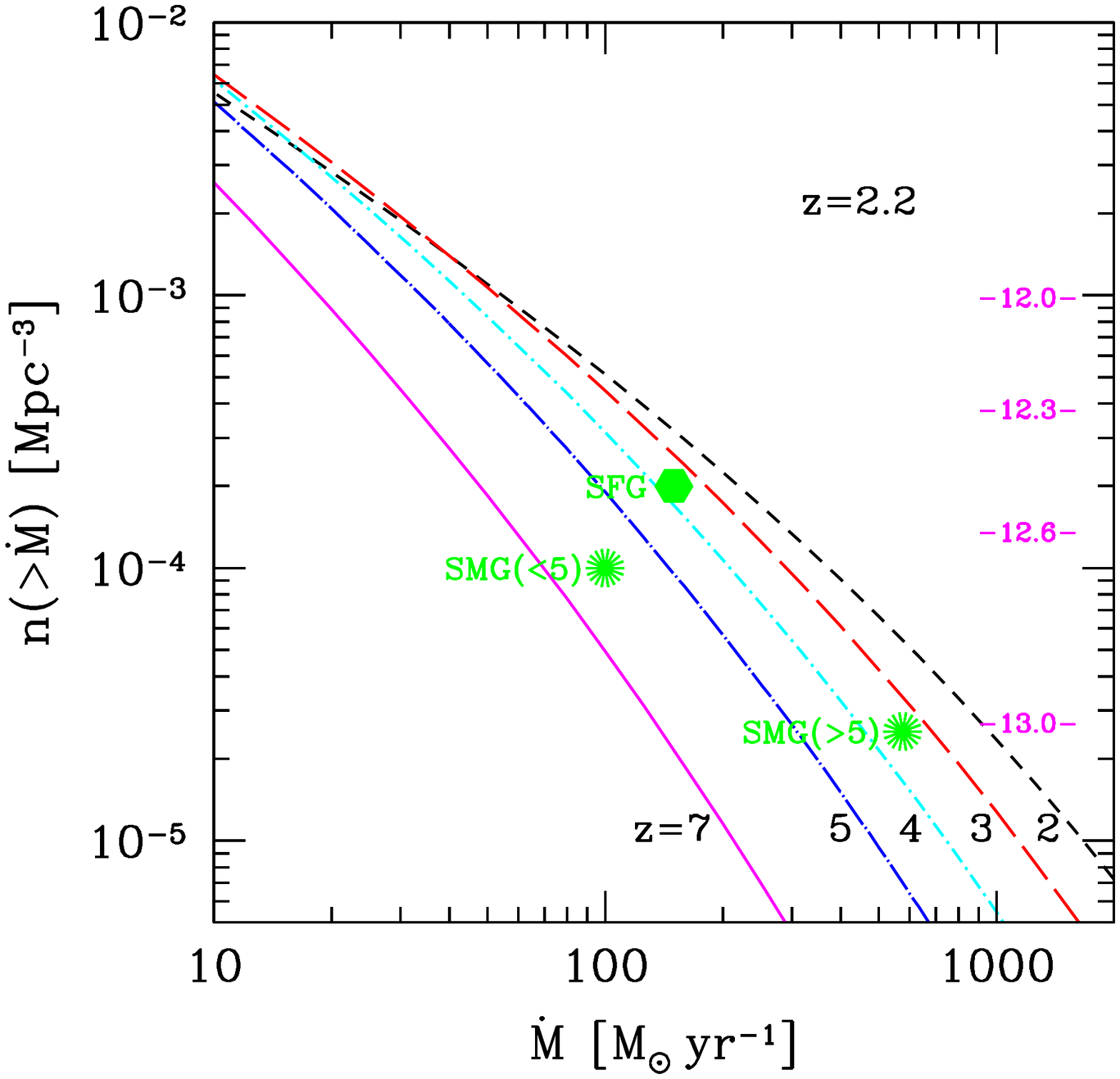}}
\vskip 0.2cm
\caption{
Comoving number density of galaxies with total gas inflow rate
higher than $\dot{M}$ at $z=2.2$, as in Fig.~4 of the main body of the Letter.
The numbers in magenta next to the right axis 
refer to $\log \Mv$ of haloes with the corresponding abundance.
{\bf Left:} Dependence on the maximum halo mass that contributes to cold
streams, for $\mst\!=\!10^{12}, 10^{12.5}, 10^{13}, 10^{14}\msun$.
{\bf Right:} Variation with redshift, $z\!=\!2,3,4,5,7$. 
}
\label{fig:dependence}
\end{figure}

As described above, the conditional probability
distribution $P(\dot{M}|\Mv)$ has been estimated by scaling the
results from the simulated haloes of $\Mv\!=\!10^{12}\msun$. 
Preliminary analysis of more massive haloes at that redshift 
(T. Goerdt et al., in preparation) indicates that the
actual inflow rate starts dropping below the adopted estimate in haloes more
massive than $\mst\!\sim\!10^{13}\msun$. 
For a first crude estimate of the effect this might have on our results 
shown in Fig.~4 of the Letter, we re-compute the comoving
number density $n(>\!\dot{M})$ as described in the main text,
but now limit the halo mass range that contributes to $\dot{M}$ 
by an upper cutoff at $\mst$. 
\Fig{dependence} shows the results for different values of $\mst$.
We see that a cutoff at $\mst\!=\!10^{13}\msun$ makes only a small
difference to $n(>\!\dot{M})$, by a factor of $\sim 2$ at the high-$\dot{M}$
regime corresponding to the bright SMGs. Thus, the decay of cold streams
above $10^{13}\msun$ is not expected to alter our results in a qualitative way.
On the other hand, we learn from the fact that the symbols for SFGs and bright
SMGs lie far above the lower curve that the high-SFR objects
at these redshifts are dominated by central galaxies in haloes more massive
than $10^{12}\msun$. In fact, we read from the figure that some of the SFGs and 
many of the bright SMGs are associated with haloes more massive than 
$3\times 10^{12}\msun$.

\Fig{dependence} also shows the predicted abundance $n(>\!\dot{M})$ at 
different redshifts, now applying no finite upper mass cutoff $\mst$.
This is justified for $z\!>\!2$ based on our preliminary investigation
of the MareNostrum galaxies at different redshifts and masses
and consistent with the conjecture of DB06\cite{db06} shown in \fig{coolzmf}.
We see that the comoving abundance of galaxies with $\dot{M}\!\sim\!150\sy$
is predicted to vary by a factor less than two between $z\!=\!2$ and 4.
By $z\!\sim\!7$ that abundance drops by an order of magnitude.
The variation with redshift is somewhat larger at the high-flux end,
toward $\dot{M}\!\sim\!10^3\sy$. 
At lower redshifts, the contribution of streams in massive haloes
above $\msh$ is most likely overestimated by this procedure, so a similar
analysis in the low-$z$ regime should impose an upper limit at
$\mst\!\simeq\!\msh$.



\end{document}